\documentclass[11pt]{article}

\usepackage[nosort]{cite}
\usepackage{color} 
\usepackage[bulletsep]{collref}
\usepackage{graphicx}
\usepackage{bbm}
\usepackage{amsmath}
\usepackage{amssymb}
\usepackage{array}
\usepackage{hyperref}

\makeatletter \@addtoreset{equation}{section} \makeatother

\makeatletter
\let\old@startsection=\@startsection
\let\oldl@section=\l@section
\renewcommand{\@startsection}[6]{\old@startsection{#1}{#2}{#3}{#4}{#5}{#6\mathversion{bold}}}
\renewcommand{\l@section}[2]{\oldl@section{\mathversion{bold}#1}{#2}}
\makeatother

\makeatletter
\let\old@makecaption=\@makecaption
\def\@makecaption{\small\old@makecaption}
\makeatother

\let\oldPhi=\Phi
\let\oldPsi=\Psi
\let\oldGamma=\Gamma
\let\oldDelta=\Delta
\let\oldSigma=\Sigma
\let\oldTheta=\Theta
\let\oldPi=\Pi
\let\oldUpsilon=\Upsilon
\renewcommand{\Phi}{\mathnormal{\oldPhi}}
\renewcommand{\Psi}{\mathnormal{\oldPsi}}
\renewcommand{\Gamma}{\mathnormal{\oldGamma}}
\renewcommand{\Sigma}{\mathnormal{\oldSigma}}
\renewcommand{\Delta}{\mathnormal{\oldDelta}}
\renewcommand{\Theta}{\mathnormal{\oldTheta}}
\renewcommand{\Pi}{\mathnormal{\oldPi}}
\renewcommand{\Upsilon}{\mathnormal{\oldUpsilon}}

\newcommand{\Action}{\mathcal{S}}
\newcommand{\Lagr}{\mathcal{L}}
\newcommand{\Ham}{\mathcal{H}}

\newcommand{\diag}{\mathop{\mathrm{diag}}}

\newcommand{\order}{\mathcal{O}}

\newcommand{\trans}{{\scriptscriptstyle\mathrm{T}}}

\newcommand{\Reals}{\mathbbm{R}}
\newcommand{\Complex}{\mathbbm{C}}
\newcommand{\Sphere}{S}  
\newcommand{\AdS}{\mathrm{AdS}}

\ifx\genfrac\sdflkaj

\else

\fi
\newcommand{\sfrac}[2]{{\textstyle\frac{#1}{#2}}}
\newcommand{\half}{\sfrac{1}{2}}

\newcommand{\Half}{\frac{1}{2}}

\newcommand{\matr}[2]{\left(\begin{array}{#1}#2\end{array}\right)}

\newcommand{\grp}[1]{\mathrm{#1}}

\newcommand{\grSO}{\grp{SO}}

\newcommand{\lrbrk}[1]{\left(#1\right)}
\newcommand{\bigbrk}[1]{\bigl(#1\bigr)}
\newcommand{\Bigbrk}[1]{\Bigl(#1\Bigr)}
\newcommand{\biggbrk}[1]{\biggl(#1\biggr)}

\newcommand{\lrsbrk}[1]{\left[#1\right]}
\newcommand{\bigsbrk}[1]{\bigl[#1\bigr]}
\newcommand{\Bigsbrk}[1]{\Bigl[#1\Bigr]}
\newcommand{\biggsbrk}[1]{\biggl[#1\biggr]}

\newcommand{\vev}[1]{\langle#1\rangle}

\newcommand{\abs}[1]{{|#1|}}

\newcommand{\nn}{\nonumber}
\newcommand{\nln}{\nonumber\\}
\newcommand{\nl}[1][0pt]{\nonumber\\[#1]&\hspace{-4\arraycolsep}&\mathord{}}

\newcommand{\earel}[1]{\mathrel{}&\hspace{-2\arraycolsep}#1\hspace{-2\arraycolsep}&\mathrel{}}
\newcommand{\eq}{\earel{=}}

\def\[{\begin{equation}}
\def\]{\end{equation}}

\makeatletter
\def\mr@ignsp#1 {\ifx\:#1\@empty\else #1\expandafter\mr@ignsp\fi}%
\newcommand{\multiref}[1]{\begingroup
\xdef\mr@no@sparg{\expandafter\mr@ignsp#1 \: }%
\def\mr@comma{}%
\@for\mr@refs:=\mr@no@sparg\do{\mr@comma\def\mr@comma{,}\ref{\mr@refs}}%
\endgroup}
\makeatother

\newcommand{\hypref}[2]{\ifx\href\asklfhas #2\else\href{#1}{#2}\fi}

\newcommand{\secref}[1]{Sec.~\multiref{#1}}

\newcommand{\appref}[1]{App.~\multiref{#1}}

\newcommand{\figref}[1]{Fig.~\multiref{#1}}
\renewcommand{\eqref}[1]{(\multiref{#1})}

\ifx\href\asklfhas\newcommand{\href}[2]{#2}\fi

\newcommand{\comma}{\quad,\quad}
\newcommand{\unit}{\mathbbm{1}}

\newcommand{\eps}{\varepsilon}

\newcommand{\be}{\begin{eqnarray}}
\newcommand{\ee}{\end{eqnarray}}

\DeclareMathOperator{\arccosh}{arccosh}
\DeclareMathOperator{\arctanh}{arctanh}

\newcommand{\cl}{{\mathrm{cl}}}
\newcommand{\nt}{{\mathrm{int}}}

\begin{document}

\thispagestyle{empty}
\begin{flushright}\footnotesize
\texttt{arXiv:1106.0495}\\
\texttt{UUITP-16/11}\\
\texttt{AEI-2011-031}%
\end{flushright}
\vspace{1cm}

\begin{center}%
{\Large\textbf{\mathversion{bold}%
A light-cone approach to  three-point functions in AdS$_5$$\times$S$^5$
}\par}

\vspace{1.5cm}

\textrm{Thomas Klose$^{a}$ and Tristan McLoughlin$^{b}$} \vspace{8mm} \\
\textit{%
$^a$ Department of Physics and Astronomy, Uppsala University \\
SE-75108 Uppsala, Sweden \\
$^b$ Max-Planck-Institut f\"ur Gravitationsphysik, Albert-Einstein-Institut, \\
Am M\"uhlenberg 1, D-14476 Potsdam, Germany
} \\
\texttt{\\ thomas.klose@physics.uu.se, tmclough@aei.mpg.de}

\par\vspace{14mm}

\textbf{Abstract} \vspace{5mm}

\begin{minipage}{14cm}
We consider worldsheet correlation functions for strings in   AdS$_5$$\times$S$^5$ using a
light-cone gauge for the worldsheet theory. We compute the saddle-point 
approximation to three-point functions of BMN vertex operators, all with large charges, by 
explicitly finding the intersection of three euclidean BMN strings. We repeat this calculation for non-BPS circular winding strings extended along a great circle of the S$^5$, though in this case the appropriate 
form of the vertex operator is uncertain. Furthermore,
we compute the spectrum of fluctuations about euclidean BMN strings for generic boundary conditions, and show that the spectrum
depends only on the total charge and not the details of the string configuration. We extend our considerations to include near-BMN 
vertex operators and through the evaluation of the string path integral make contact with the light-cone string field theory calculations of gauge theory three-point structure constants.
\end{minipage}

\end{center}

\newpage

\section{Introduction}

The calculation of worldsheet correlation functions of vertex operators for strings in AdS$_5\times$S$^5$ is, by the AdS/CFT conjecture \cite{Maldacena:1997re}, equivalent to the computation of space-time correlation functions in the boundary theory. While there has been a tremendous amount of work on this correspondence, for the most part explicit holographic calculations of correlators have been restricted to BPS operators and to the supergravity limit, for example \cite{D'Hoker:1998tz, D'Hoker:1999ea,Lee:1998bxa}. A key difficulty in going beyond the supergravity approximation is the identification of  the appropriate string vertex operator corresponding to a given gauge invariant operator in the boundary theory. Not knowing the exact vertex operators nor how to exactly quantize the worldsheet theory, it is useful to take a semiclassical  approach and consider states with charges that scale like the string worldsheet coupling $\sqrt{\lambda}\gg1$ \cite{Polyakov:2001af,Tseytlin:2003ac}. In this case the string path integral for the correlation functions can be evaluated in the saddle-point approximation. Recently, there has been renewed interest in the identification of the semiclassical vertex operators \cite{Buchbinder:2010gg, Buchbinder:2010vw, Roiban:2010fe} and the calculation of their Euclidean signature two-point functions. Worldsheet correlation functions in Lorentzian signature have also been recently studied in  \cite{Janik:2010gc} where, moreover, it was shown how the strong coupling calculation reproduces the correct space-time dependence of the gauge theory correlators. 

In this semiclassical limit one can further attempt to calculate three-point functions, however now in addition to identifying the correct vertex operators one must find the classical solution which provides the appropriate saddle point. Recent progress \cite{Zarembo:2010rr,Costa:2010rz} has involved studying three-point functions where two operators are ``heavy'' and have charges which  scale as $\sqrt{\lambda}$ and a third ``light'' operator which has charges that are constant or scale as $\lambda^{1/4}$ \cite{Roiban:2010fe,Hernandez:2010tg, Ryang:2010bn, Georgiou:2010an,Russo:2010bt,Hernandez:2011up}. In this case the saddle-point surface is  just that sourced by the two heavy operators and the three-point function can be found by evaluating the ``light'' vertex operator on this classical surface. Similar considerations have been extended to four-point functions \cite{Buchbinder:2010ek, Arnaudov:2011wq}, open strings \cite{Bak:2011yy}, giant magnons \cite{Park:2010vs}, giant gravitons \cite{Bissi:2011dc}, dyonic strings \cite{Bai:2011su}, Wilson loops \cite{Zarembo:2002ph, Alday:2011pf} and even to include finite-size effects \cite{Ahn:2011zg,Lee:2011fe}. Nonetheless, a complete understanding of the exact form of the vertex operators and the appropriate saddle-point surfaces for correlators of three operators with equally large charges remains a challenging problem. Furthermore, at strong coupling there remains much to be done in going beyond the semiclassical approximation and including quantum corrections. 

One limit in which the match between string vertices and gauge theory operators is better understood and where quantum corrections have been calculated is the plane-wave limit. This limit \cite{Berenstein:2002jq} can be understood as taking a BMN string, a point-like string sitting at the center of the AdS space and rotating along a great circle of the sphere, as the string vacuum. The string worldsheet theory is exactly solvable \cite{Metsaev:2001bj, Metsaev:2002re} in a light-cone gauge adapted to these geodesics  and it is possible to make a match between the string states and so-called BMN operators in the gauge theory (see \cite{Beisert:2002tn} for a useful definition of these operators). In this limit it was also possible to construct the cubic Hamiltonian of light-cone string field theory which describes the splitting and joining of strings
(see \cite{Spradlin:2003xc, Russo:2004kr} for reviews). Furthermore, this vacuum played an important 
calculational and conceptual role in studies of integrability in the planar limit of the AdS/CFT duality (for reviews
see \cite{Arutyunov:2009ga}  and \cite{Beisert:2010jr}\footnote{For discussion of the plane-wave limit  and its quantization see
particularly chapters II.1, II.2 and II.3 \cite{Tseytlin:2010jv, McLoughlin:2010jw,Magro:2010jx}.}). The integrability 
of the worldsheet theory allowed for the exact solution to the spectrum of string energies, correspondingly 
gauge theory anomalous dimension, and thus  gauge theory two-point functions. It may hopefully lead to a greater understanding of the matching between string vertex operators and gauge theory operators. It has already been shown that  integrable methods, 
e.g. the algebraic Bethe ansatz for the spin-chain description of operators, can be useful in the calculation of three-point functions at weak coupling \cite{Okuyama:2004bd, Roiban:2004va, Alday:2005nd, Escobedo:2010xs} and even matched to strong coupling results 
\cite{Escobedo:2011xw}. 

In this work we consider the strong coupling description of euclidean BMN strings as saddle points of the path integral for three-point correlators and the quantization about this classical approximation. We make use of a  light-cone gauge which starts from the Poincar\'e coordinates for AdS$_5$ and forms the light-cone directions from two boundary coordinates \cite{Metsaev:2000yu,Tseytlin:2002gz}. In the literature this light-cone gauge has been used somewhat less frequently, however recently it has proven very useful in efficiently calculating the quantum corrections to various semiclassical string configurations \cite{Giombi:2009gd, Giombi:2010fa,Giombi:2010zi,Giombi:2010bj}. In this gauge, point-like solutions are particularly simple being merely straight lines. 
Starting with the two-point functions, where the string topology is a cylinder, we interpret the chiral primary vertex operators given in \cite{Buchbinder:2010ek} (following on from \cite{Berenstein:1998ij, Zarembo:2010rr}) as contributing boundary terms to the string action, which determine the string state at the initial and final times. We show that the analytic continuation of the point-like  solutions satisfies all the appropriate equations of motion, including those at the boundary. 
This slightly differs conceptually from the calculation in e.g. \cite{Buchbinder:2010gg, Buchbinder:2010vw, Roiban:2010fe}, which consider the worldsheet as a plane with vertex operators inserted at specific points, though  of course the two prescriptions should be equivalent by the usual 
state/operator conformal mapping. In the cases considered in this work, as the vertices evaluated on the solution do not depend on the worldsheet coordinate,  the mapping is essentially trivial however for more general solutions this may need to be treated
differently. Evaluating the action on this solution reproduces the space-time dependence of Euclidean two-point functions in conformal field theory for chiral primary operators. In this, we are simply recasting the results of \cite{Janik:2010gc, Buchbinder:2010gg, Buchbinder:2010vw, Roiban:2010fe}  into light-cone gauge. However, we can then go beyond the leading result and include  corrections from quantum fluctuations about this result, moreover one can define the vertex operators for BMN operators with a few added impurities. Much of this closely parallels the calculation of string energies for BMN strings and one key point is that the spectrum for the general class of solutions does not depend on the specific boundary conditions or string orientation on the sphere but only the total charge. 

The same euclidean BMN strings can be used to find the classical solution sourced by three BMN vertex operators. We
follow the calculation of \cite{Janik:2010gc} in finding the intersection of three BMN strings and then minimizing the action
by varying the intersection point. We are able to explicitly solve this minimization problem and so find the complete solution. 
As has already been shown, though without finding the explicit solution, evaluating the action on this saddle
point gives a holographic derivation of the space-time dependence for gauge theory three-point functions. 
Having the explicit form of the solution allows us to extend the results from the two-point functions and to calculate the fluctuations
about the three-point functions. In this, we are able to make contact with the results of light-cone string field theory, though with a few caveats. Moreover, our considerations are valid not only for extremal correlators but also for non-extremal ones.

There have been several earlier works analyzing the holographic calculation of the three-point functions of BMN operators, of particular note are those based on the GKP-Witten \cite{Gubser:1998bc, Witten:1998qj} definition of the AdS/CFT duality \cite{Dobashi:2002ar,Yoneya:2003mu,Dobashi:2004nm,Lee:2004cq,Shimada:2004sw}. These calculations lift the supergravity calculation of three-point functions to include string effects  by  making use of the light-cone string field theory results. In many respects our calculation of the quantum fluctuations simply reproduces these results although within a slightly different framework that makes contact with more recent progress regarding three-point functions of far from BPS semiclassical strings. 
 
In particular we are able to extend our results, at least for the leading semiclassical contribution, to circular winding strings. These solutions were found in \cite{Frolov:2003qc}, with the general class of solutions being described in \cite{Arutyunov:2003za}. 
Specifically, we are able to show that by gluing three segments together, we can find a saddle-point solution which has the boundary conditions appropriate to the simplest circular winding strings at all three boundary points. In \cite{Ryang:2010bn} a proposal for the vertex operators corresponding to the circular winding strings was made. While this vertex operator does indeed source the saddle-point surface corresponding to the two-point correlator this does not guarantee that it is correct. For one, there could be subleading polynomial terms, which though they will not affect the saddle-point calculation of the surface will modify the strong coupling prediction for the correlators. Further, it is not clear that the boundary conditions are even sufficient to uniquely determine the exponentially large contributions and in the earlier work \cite{Buchbinder:2010gg} a different
proposal was made which involved T-dualized angles. This vertex also provides appropriate boundary conditions 
but, in our formulation,  only after one changes the boundary conditions for the bulk string action. Nonetheless, we evaluate the action, including the boundary terms corresponding to the vertices of \cite{Ryang:2010bn},  which thus provides a strong coupling approximation to the three point correlators of these vertices. The result using the vertices of \cite{Buchbinder:2010gg} is essentially the same. 

\section{Coordinates and geodesics}

In this paper, we work with \emph{Euclidean} $\AdS_5$, defined as the surface $X_0^2+\ldots +X_4^2 - X_5^2 = -R^2$ embedded in $\Reals^{5,1}$ with metric $(+,\ldots,+,-)$. This surface can either be parametrized by global coordinates ($i=1,\ldots,4$)
\be
  X_0 = R \cosh\rho \, \sinh t
  \comma
  X_i = R \sinh\rho \, \Omega_i
  \comma
  X_5 = R \cosh\rho \, \cosh t
  \; ,
\ee
where $\Omega_i$ is a unit vector, or by Poincar\'e coordinates
\be
  \vec{X} = R \, \frac{\vec{x}}{z}
  \comma
  X_4 = \frac{R}{2z} \lrbrk{-1 + z^2 + \vec{x}^{\,2} }
  \comma
  X_5 = \frac{R}{2z} \lrbrk{ 1 + z^2 + \vec{x}^{\,2} }
  \; .
\ee
where we have introduced the vector notation $\vec{x} = (x_0,x_1,x_2,x_3)$ for the coordinates on the boundary of AdS. Unlike in the case of Lorentzian signature, the Poincar\'e coordinates cover the entire Euclidean AdS space.

\paragraph{Geodesics.} In Poincar\'e coordinates, the geodesics are semi-circles with center at the boundary $z=0$. These geodesics are the Wick rotation of light-like geodesics in Lorentzian AdS. Explicitly, the geodesic that starts at point $(\vec{x},z) = (\vec{a}_1,0)$ and ends at point $(\vec{a}_2,0)$ can be parametrized by
\be
\label{eq:poin_geod}
  \vec{x}(\tau) = \frac{\vec{a}_2 - \vec{a}_1}{2} \tanh \kappa \tau + \frac{\vec{a}_1 + \vec{a}_2}{2}
  \comma
  z(\tau) = \frac{\abs{\vec{a}_2 - \vec{a}_1}}{2\cosh \kappa \tau}
  \comma
  -\infty \leq \tau \leq \infty
\ee
and satisfies
\be
  \lrbrk{\vec{x} - \frac{\vec{a}_1 + \vec{a}_2}{2}}^2 + z^2 = \lrbrk{\frac{\vec{a}_2 - \vec{a}_1}{2}}^2 \; .
\ee
In global coordinates, this geodesic becomes
\be
  \tanh t(\tau)      \eq \frac{\abs{\vec{a}_2-\vec{a}_1} \tanh\kappa\tau}{1+\half (\vec{a}_1^{\,2}+\vec{a}_2^{\,2})+ \half (\vec{a}_2^{\,2}-\vec{a}_1^{\,2}) \tanh\kappa\tau} \\
  \cosh^2 \rho(\tau) \eq \frac{\cosh^2\kappa\tau}{(\vec{a}_2-\vec{a}_1)^2} \lrsbrk{ (1+\half (\vec{a}_1^{\,2}+\vec{a}_2^{\,2})+ \half (\vec{a}_2^{\,2}-\vec{a}_1^{\,2}) \tanh\kappa\tau)^2 - (\vec{a}_2-\vec{a}_1)^2 \tanh^2\kappa\tau }\nn\\
\ee
and $\Omega_{1} = \Omega_{2} = \Omega_{3} = 0$, $\Omega_{4} = 1$.

\begin{figure}%
\begin{center}
\includegraphics[scale=1]{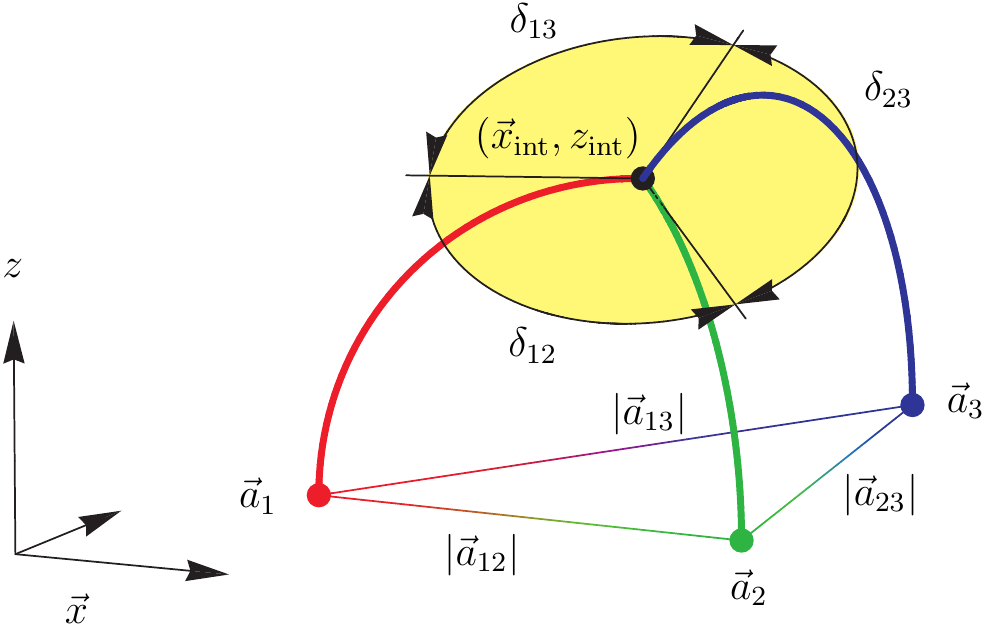}
\end{center}
\caption{\textbf{Intersecting geodesics.} Each segment is a geodesic and the location of the intersection point is found by demanding that the overall proper length is minimal, where, however, the length of each segment is weighted by the conformal dimension that is associated with the corresponding string.}%
\label{fig:strings3d}%
\end{figure}

\paragraph{Intersecting geodesics.} In the discussion of the string three-point functions, we will arrive at a configuration where three geodesics of the form \eqref{eq:poin_geod} intersect in the bulk at some point $(\vec{x}_\nt,z_\nt)$, see \figref{fig:strings3d}. For given locations $\vec{a}_i$, where the three geodesics reach the boundary, the intersection point is determined by extremizing\footnote{In this case, it is maximizing.} the function\cite{Janik:2010gc}
\be \label{eqn:minimize-this}
  \mathcal{B} = \sum_{i=1}^3 \Delta_i \ln\frac{z_\nt}{z_\nt^2 + (\vec{x}_\nt-\vec{a}_i)^2} \; .
\ee
In our computation in \secref{sec:class-3pt}, we will obtain \eqref{eqn:minimize-this} as the boundary action which encodes the specifics of the interacting strings, e.g.\ the dimensions $\Delta_i$ of the operators to which they are dual. Geometrically, \eqref{eqn:minimize-this} the total proper length of the three interacting strings  where each segment is weighted by the corresponding dimension. This is seen by expanding the geodesic distance between $(\vec{a}_i,\eps)$ and $(\vec{x}_\nt,z_\nt)$ for $\eps\to 0$:
\be
  \arccosh\lrbrk{1+\frac{(\vec{x}_\nt-\vec{a}_i)^2+ (z_\nt-\eps)^2}{2z_\nt \eps}} = -\ln\frac{z_\nt \eps}{z_\nt^2 + (\vec{x}_\nt-\vec{a}_i)^2} + \order(\eps) \; .
\ee

Finding the intersection point analytically is greatly facilitated by the introduction of the variables\footnote{This was suggested to us by Joe Minahan who first obtained the result \protect\eqref{eqn:intersection-point}, and furthermore interpreted the extremization conditions as the conservation laws for the canonical momenta of the string at the intersection \cite{Minahan:2011unpub}.}
\be
\label{eq:alphadef}
  \alpha_1 = \Delta_2 + \Delta_3 - \Delta_1
  \qquad \mbox{and cyclic permutations of $1,2,3$} \; .
\ee
In terms of these quantities, the coordinates of the intersection point are given by
\be \label{eqn:intersection-point}
 \vec{ x}_\nt \eq \frac{\alpha_2 \alpha_3~ \vec{a}_{23}^{\,2} \, \vec{a}_{1} + \alpha_1 \alpha_3~ \vec{a}_{13}^{\,2}  \, \vec{a}_{2} + \alpha_1 \alpha_2~ \vec{a}_{12}^{\,2} \, \vec{a}_{3}}{\alpha_2 \alpha_3~ \vec{a}_{23}^{\,2} + \alpha_1 \alpha_3~ \vec{a}_{13}^{\,2} + \alpha_1 \alpha_2 ~\vec{a}_{12}^{\,2}} \\[3mm]
  z_\nt \eq \frac{\sqrt{\alpha_1 \alpha_2 \alpha_3 (\alpha_1 + \alpha_2 + \alpha_3) } \, \abs{\vec{a}_{23}} \abs{\vec{a}_{13}} \abs{\vec{a}_{12}}}{\alpha_2 \alpha_3~ \vec{a}_{23}^{\,2} + \alpha_1 \alpha_3~ \vec{a}_{13}^{\,2} + \alpha_1 \alpha_2~ \vec{a}_{12}^{\,2}} \nn
\ee
with $\vec{a}_{ij}=\vec{a}_i-\vec{a}_j$. This result is physically sensible only for $z_\nt\ge0$, i.e.\ all $\alpha$'s must be positive. This imposes the triangle inequality
\be
  \abs{\Delta_1 - \Delta_2} \le \Delta_3 \le \Delta_1 + \Delta_2   \qquad \mbox{and cyclic} \; ,
\ee
on the conformal dimensions. The tangents to the three segments at the intersection point lie in the same plane, see \figref{fig:strings3d}, and the angles $\delta_{ij}$ between the segments $i$ and $j$ are given by
\be
  \cos\delta_{12} = \frac{\Delta_3^2 - \Delta_1^2 - \Delta_2^2}{2 \Delta_1 \Delta_2}   \qquad \mbox{and cyclic} \; .
\ee
If all dimensions $\Delta_i$ are equal to each other, then all angles are equal to $\delta_{ij}=120^\circ$. If one dimension is equal to the sum of the other two, say $\Delta_3 = \Delta_1 + \Delta_2$ so that $\alpha_3=0$, then segment ``3'' shrinks to zero length and the intersection point coincides with the point $\vec{a}_3$ on the boundary where the other two segments arrive with parallel tangents ($\delta_{12} = 0$, $\delta_{13} = \delta_{23} = 180^\circ$), thus in this case the string is essentially the product
of two two-point correlators. This corresponds to an extremal correlator. It is interesting to note that the limit, $\alpha_3\to0$, is smooth and in fact one can define the extremal correlator as the analytic continuation of the non-extremal version as 
suggested in \cite{Liu:1999kg}
\footnote{A similar calculation was performed by Buchbinder and Tseytlin
 \cite{Buchbinder:2011unpub}
who considered three CPO
operators and assumed that once fermions are included the  correlator
 localizes to just the usual supergravity expression in terms of three
AdS propagators which essentially gives \eqref{eqn:minimize-this}. Then, by performing a stationary point approximation, they 
were also able to show that the there is no non-trivial trajectory for $\alpha_3=0$
but that it should rather be defined by analytic continuation of the nonextremal case.}.

We can also reproduce the ``heavy-heavy-light'' configuration which is discussed so extensively in the recent literature by first setting two dimensions equal to each other, say $\Delta_1=\Delta_2$, and then the third to zero. Then the angles are given by $\delta_{12} = 180^\circ$ and $\delta_{13} = \delta_{23} = 90^\circ$, i.e.\ the heavy segments ``1'' and ``2'' form a semi-circle to which the light segment ``3'' is attached without being able to pull the intersection point toward $\vec{a}_3$. 

\paragraph{$\AdS_5\times\Sphere^5$ coordinates and complexification.} We parametrize points on $\Sphere^5$ by a unit vector $\mathbf{u}$ in $\Reals^6$, where we use bold-face in order to distinguish these vectors from the 4-vector $\vec{x}$ that parametrizes points on the AdS-boundary. In these coordinates, the metric on $\AdS_5\times\Sphere^5$ reads
\be
\label{eq:Poincare_coords}
  ds^2=z^{-2}\bigbrk{d\vec{x}\,^2 + dz^2}+d\mathbf{u}^2 \; .
\ee
It is convenient to introduce an unconstrained 6-vector $\mathbf{z} = z \mathbf{u}$ which mixes the radial part of AdS with the 5-sphere. In terms of this vector, the metric becomes simply
\be
  ds^2 = \mathbf{z}^{-2} \bigbrk{ d\vec{x}\,^2 + d\mathbf{z}^2 } \; .
\ee

Later, we will compute the saddle points of the path integral for strings propagating in this background. Such saddle points are in general complex. Therefore, we allow $\vec{x} \in \Complex^4$ and $\mathbf{z} = \Complex^6$. However, we will retain the definition of the norm to be $\abs{\mathbf{z}} = \sqrt{\mathbf{z}\cdot\mathbf{z}}$ and do \emph{not} use $\sqrt{\mathbf{z}\cdot\mathbf{z}^*}$.

\section{Vertex operators}

The structure of vertex operators corresponding to semiclassical string states in AdS was described in
\cite{Polyakov:2001af,Tseytlin:2003ac}. A string with charges $\{Q_i\}$, for example string energy  (equivalently conformal dimension) $\Delta$,  AdS spin $S$, or  angular momentum on the sphere, $J$,\footnote{The string will also in general depend on discrete quantum numbers such as winding or mode numbers.} 
 is created at the location $\vec{a}$ on the boundary of AdS by an integrated vertex operator $V_{\{Q_i\}}(\vec{a})$. It is given by an integral over the worldsheet,
\be
  V_{\{Q_i\}}(\vec{a}) = \int\!\frac{d\sigma\,d\tau}{2\pi} \: V_{\{Q_i\}}(\sigma,\tau,\vec{a}) \; ,
\ee
where the unintegrated vertex $V_{\{Q_i\}}(\sigma,\tau,\vec{a})$ 
is a function of the target space bosonic and fermionic coordinates 
and their derivatives and thus implicitly depends on the worldsheet coordinates. 
Furthermore, the vertex operator generically decomposes into a part $W$ which scales exponentially in the charges e.g.  
$(\dots)^{Q_i}$ and a polynomial part $U$:
\be
  V_{\{Q_i\}}(\sigma,\tau,\vec{a}) = W_{\{Q_i\}}(\sigma,\tau,\vec{a}) U(\sigma,\tau) \; .
\ee
In the large charge limit, when the charges scale like $Q_i\sim \sqrt{\lambda}$, the exponential part can be interpreted as
providing a boundary action for the path integral, it thus acts as a source for the saddle-point worldsheet in the semiclassical approximation. The polynomial part generically involves derivative terms and fermions, which can encode information such as the mode number and polarization of the excited string state.

\paragraph{BMN strings.} The exponential part of the non-integrated vertex operator for (near-)BMN strings is given by \cite{Polyakov:2001af,Tseytlin:2003ac}
\be \label{eqn:def-genvertop}
  W^{\mathrm{BMN}}_{\Delta,J}(\sigma,\tau,\vec{a},\mathbf{n}) = \biggbrk{\frac{ \abs{\mathbf{z}} }{ \mathbf{z}^2 + (\vec{x} - \vec{a})^2}}^\Delta
    \biggbrk{\frac{ \mathbf{n}\cdot \mathbf{z} }{ \abs{\mathbf{z}} }}^J \; .
\ee
It creates a string with angular momentum $J$ on the 5-sphere in a plane that is specified by the complex polarization 6-vector $\mathbf{n}$ which satisfies\footnote{We could have normalized $\mathbf{n}$ to unity, but that would have made \protect\eqref{eqn:recipocity} more inconvenient.}
\be \label{eqn:properties-n}
  \mathbf{n}^2 = 0
  \comma
  \mathbf{n}\cdot\mathbf{n}^* = 2 \; .
\ee
The simplest example would be $\mathbf{n} = (1,i,0,0,0,0)$. More general vertex operators would be obtained by the replacement
\be
\label{eq:vertex_harmonics}
  \biggbrk{\frac{ \mathbf{n}\cdot \mathbf{z} }{ \abs{\mathbf{z}} }}^J \longrightarrow Y(\hat{\mathbf{z}}) \; ,
\ee
where $Y(\hat{\mathbf{z}})$ is a spherical harmonic of $\grSO(6)$ and $\hat{\mathbf{z}} = \frac{\mathbf{z}}{\abs{\mathbf{z}}}$. Most conveniently, these functions are written as homogeneous polynomials
\be
  Y(\hat{\mathbf{z}}) = C_{MNO\ldots} \hat{z}^M \hat{z}^N \hat{z}^O \cdots \; .
\ee
For this to be an irreducible representation, the tensor $C_{MNO\ldots}$ has to be symmetric and completely traceless. We will stick to the special case \eqref{eqn:def-genvertop}, which corresponds to the highest weight state and is obtained by setting
\be
  C_{MNO\ldots} = n_M n_N n_O \cdots \; .
\ee 

The bosonic quadratic-in-derivatives part of the BMN vertex operator $U=U_{\mathrm{I}}+U_{\mathrm{II}}+U_{\mathrm{III}}$ as written in \cite{Buchbinder:2010ek} (based on \cite{Berenstein:1998ij, Zarembo:2010rr}) consists of the components
\be \label{eqn:BMN-vertex-op-U}
  U_{\mathrm{I}} \eq \frac{\sqrt{h}h^{ab}}{ 8\mathbf{z}^2 }\Big[(\partial_a {\vec x}\cdot \partial_b {\vec x})-(\partial_a \mathbf{z}\cdot \partial_b \mathbf{z})\Big] \; , \\
  U_{\mathrm{II}}+U_{\mathrm{III}} \eq \frac{\sqrt{h}h^{ab}}{(\mathbf{z}^2+\vec{x}^{\,2})^2}\left( \Big[\vec{x}^{\,2}(\partial_a \abs{\mathbf{z}})(\partial_b \abs{\mathbf{z}})-(\vec{x}\cdot\partial_a\vec{x})(\vec{x}\cdot\partial_b\vec{x})\Big]
+\frac{\mathbf{z}^2-\vec{x}^{\,2}}{\abs{\mathbf{z}}}(\vec{x}\cdot \partial_a\vec{x})\partial_b\abs{\mathbf{z}}\right) \; . \nn
\ee
The complete expression for $U$ will involve fermionic terms which can in principle be derived as in \cite{Zarembo:2010rr} but expanding the full superstring action rather than just the bosonic part.

\section{String action in light-cone gauge}

In this section we wish to briefly review the AdS light-cone gauge fixing, for a recent treatment see e.g. \cite{Giombi:2009gd}, 
and the corresponding calculation of the string path integral. Here we focus on the bosonic fields and discuss the fermions in \appref{app:fermions_fluc}. Starting from the $\AdS_5\times\Sphere^5$ metric in Poincar\'e coordinates
\eqref{eq:Poincare_coords}, we introduce the AdS light-cone combinations
\be
 x^{\pm}=\tfrac{1}{\sqrt{2}}(x^3\pm i x^0)
 \comma
 x=\tfrac{1}{\sqrt{2}}(x^1+i x^2)
 \comma
 \bar{x}=\tfrac{1}{\sqrt{2}}(x^1-i x^2)~.
\ee 
In these coordinates the product of two vectors is $\vec{a}\cdot\vec{b} = a^+ b^- + a^- b^+ + a \bar{b} + \bar{a} b$, where
 $a_+ = a^-$ and $a_- = a^+$ and so the  metric now reads
\be
  ds^2 = \mathbf{z}^{-2}\bigbrk{ 2dx^+ dx^- + 2dx d\bar{x} + d\mathbf{z}^2 }~.
\ee
For the classical solutions we consider, this form of light-cone gauge is particularly useful. However, we can be slightly more 
general and use the generic notation
\be
  ds^2 = G_{\mu\nu}(X) dX^\mu dX^\nu = 2 G_{+-}(X) dX^+ dX^- + G_{AB}(X) dX^A dX^B \; ,
\ee
and substitute later $X^\mu = (\vec{x},\mathbf{z})$ and $G_{\mu\nu} = \delta_{\mu\nu} / \mathbf{z}^2$. The bosonic string action is 
\be
\label{eq:com_bos_action}
\Action \eq \frac{\sqrt{\lambda}}{2\pi} \int^{2 \pi}_0 d\sigma \int_{\tau_1}^{\tau_2} d\tau \: \Lagr 
\comma
\Lagr = \frac{1}{2} \sqrt{h} \, h^{ab} \, \partial_a X^\mu \partial_b X^\nu \, G_{\mu\nu}(X) \; ,
\ee
where $h_{ab}$ is the Euclidean\footnote{This action is also good for a Lorentzian worldsheet with signature $(+,-)$, but for signature $(-,+)$, we would have to change the overall sign of the action. Moreover, for Lorentzian worldsheet of either signature, the sign of the $\acute{X}^\mu \acute{X}^\nu G_{\mu \nu}$-term in \protect\eqref{eqn:ham} below would change.} worldsheet metric and $h=\abs{\det(h_{ab})}$. We wish to calculate the worldsheet correlators defined by the usual Euclidean path integral with insertions,
\be
  \vev{\ldots} =\int \mathcal{D}X\, \mathcal{D}p \, \mathcal{D}h \: (\ldots) \, e^{-\Action[p,X,h]} \; .
\ee
The momentum densities are defined to be 
\be
p_\mu = \frac{\partial \Lagr}{\partial \dot{X}^\mu} =  \sqrt{h} \, h^{\tau b} \, \partial_b X^\nu \, G_{\mu\nu} \; ,
\ee
so we can write the Lagrangian as 
\be
  \Lagr = p_\mu \dot{X}^\mu - \Ham
\ee
with 
\be \label{eqn:ham}
  \Ham = \frac{1}{2 \sqrt{h} \, h^{\tau\tau}}\Bigbrk{ G^{\mu\nu} p_\mu p_{\nu} - \acute{X}^\mu \acute{X}^\nu G_{\mu \nu} }
  -\frac{h^{\tau\sigma}}{h^{\tau\tau}} \, \acute{X}^\mu p_\mu \; .
\ee
As is usual for diffeomorphism invariant theories the Hamiltonian is  a sum of constraints with components of  the metric acting as Lagrange multipliers. We can thus integrate out\footnote{In principle there is a non-trivial Jacobian factor from the path integral measure  and moreover if the insertions depend on $h^{ab}$ these must be treated carefully. However such terms will not 
be relevant to our considerations and for the insertions we will simply insert the appropriate gauge fixed versions.} the worldsheet metric $h^{ab}$ which results in the constraints
\be  \label{eqn:int-over-p} 
  G^{\mu\nu} p_\mu p_\nu - \acute{X}^\mu \acute{X}^\mu G_{\mu\nu} = 0
  \comma
  p_\mu \acute{X}^\mu = 0 \; .
\ee
We can further impose the light-cone gauge
\be
  X^+ = \tau
  \comma
  p_- = s \; ,
\ee
where $s$ is a constant. In this gauge, the constraints \eqref{eqn:int-over-p} become
\be
  && p_+ = - \frac{G_{+-}(X)}{2s} \Bigbrk{ G^{AB} p_A p_B - \acute{X}^A \acute{X}^B G_{AB} } \equiv -\Ham_{\mathrm{lc}}(p_A, X^A,s)
  \; , \\[1mm] &&
  s \acute{X}^- + p_A \acute{X}^A = 0 \; ,
\ee
where the first equation defines the light-cone Hamiltonian $\Ham_{\mathrm{lc}}(p_A, X^A,s)$. These can be used to  remove\footnote{An integral over the zero mode of $X^-$ is left which results in an important non-locality in the gauge fixed theory particularly
in the definition of the supercharges.  The remaining integral over $s$ imposes boundary conditions on this zero mode.} the path integrals over $p_+$ and $X^-$. Moreover, as light-cone gauge is a physical gauge the ghost contributions decouple. Thus, one is left with a path integral over the transverse coordinates and momenta in addition to an ordinary integral over $s$
\be
\vev{ \ldots } \eq \int\! \mathcal{D}X^A \, \mathcal{D}p^A \, ds \:  (\ldots) \, e^{ - \frac{\sqrt{\lambda}}{2\pi} \int\limits_{0}^{2\pi} \! d\sigma \int\limits_{\tau_1}^{\tau_2} \! d\tau \:
\Bigsbrk{p_A \dot X^A + s \dot X^- - \Ham_{\mathrm{lc}}(p_A, X^A)} 
} \; .
\ee
Leaving the general discussion and specializing to $\AdS_5\times\Sphere^5$, the path integral becomes 
\be
  \vev{\ldots} \eq \int\! \mathcal{D}x \, \mathcal{D}\bar{x} \, \mathcal{D}\mathbf{z} \, \mathcal{D}p \, \mathcal{D}\bar{p} \, \mathcal{D}\mathbf{p} \, ds \: (\ldots) e^{-\Action}
\ee
with the action
\be \label{eqn:bulk-action}
 \Action = \frac{\sqrt{\lambda}}{2\pi} \int_{0}^{2\pi}\!d\sigma \int_{\tau_1}^{\tau_2}\!d\tau\: \biggbrk{
  p \dot{\bar{x}} + \bar{p} \dot{x} + \mathbf{p} \dot{\mathbf{z}} + s \dot{x}^- - \Ham_{\mathrm{lc}} } \; ,
\ee
where the light-cone Hamiltonian is
\be
  \Ham_{\mathrm{lc}} = \frac{1}{s} \lrbrk{ p \bar p + \Half \mathbf{p}^2
 -\frac{\acute{x} \acute{\bar{x}} }{\mathbf{z}^4 } - \frac{\acute{\mathbf{z}}^2}{2 \mathbf{z}^4 } } \; .
\ee
Thus we find an effective path integral for the physical degrees of freedom. This action is essentially equivalent 
to that found by directly imposing the diagonal light-cone gauge, $h^{ab}=\diag(\mathbf{z}^2,\mathbf{z}^{-2})$, on the Lagrangian, dropping the $x^-$ 
degree of freedom and analytically continuing to Euclidean signature, see for example \cite{Giombi:2009gd}. This also provides a convenient method for finding the action for the fermionic fields which is necessary when we wish to perform the fluctuation 
analysis about leading classical saddle points. It is to the
determination of such classical configurations that we now turn our attention.

\section{Classical two-point function}
\label{sec:class-2pt}

We consider the two-point correlator of two BMN vertex operators located at the boundary positions $\vec{a}_1$ and $\vec{a}_2$ for worldsheet times\footnote{In light-cone gauge, the boundary locations of the vertex operators and their
worldsheet times are obviously related by $x^+=\tau$.} $\tau_1$, $\tau_2$ and rotating in planes intersecting the S$^5$ described by $\mathbf{n}_1$, $\mathbf{n}_2$,
\be
  \vev{ V_1(\tau_1,\vec{a}_1,\mathbf{n}_1) V_2(\tau_2,\vec{a}_2,\mathbf{n}_2) } \; .
\ee
For the most part we are simply recasting the results from conformal gauge calculations \cite{Janik:2010gc, Buchbinder:2010gg, Buchbinder:2010vw} into light-cone gauge. However, this is useful for introducing the notation describing the saddle-point configurations (Euclidean classical solutions) and to highlight the differences, most notably the absence of a marginality condition for the vertex operators. 

The exponential parts of the vertex operators, which scale as $\sqrt{\lambda}$, supply a boundary action $\mathcal{B}$ for the path integral. Then we are left with the expectation value of the polynomial parts of the vertex operators 
\be
  \vev{ V_1(\tau_1,\vec{a}_1,\mathbf{n}_1) V_2(\tau_2,\vec{a}_2,\mathbf{n}_2) }_{\Action} =
  \vev{ U_1(\tau_1) U_2(\tau_2) }_{\Action+\mathcal{B}(W_1,W_2)} ~.
\ee
The bulk action $\Action$ is given in \eqref{eqn:bulk-action} while the boundary action is
a sum over contributions from the different boundaries associated with the individual vertex operator insertions or explicitly for two-point functions, $\mathcal{B} = \mathcal{B}_1 + \mathcal{B}_2$, with 
\be
\label{eq:boundary_act}
 \mathcal{B}_i = \frac{\sqrt{\lambda}}{2\pi} \int_{0}^{2\pi}\!d\sigma \int_{\tau_1}^{\tau_2}\!d\tau\: \biggbrk{
  - \frac{1}{\sqrt{\lambda}} \ln W_i(\tau,\vec{a}_i,\mathbf{n}_i) \delta(\tau-\tau_i)
  } \; .
\ee
Here we take the definition of the integrated vertex operators to be $\exp\bigsbrk{\frac{1}{2\pi} \int d\sigma \ln V_{i}(\tau_i,{\sigma},\vec{a})}$ which is natural when interpreting the exponentially large part of the vertex as part of an action. The standard definition, taking the integral inside the logarithm, gives the same answer as, on the solution, the vertex is independent of $\sigma$. This most likely will not be true for more general solutions which are $\sigma$ dependent. In this section we evaluate the action (bulk and boundary) at the saddle point while in the next section we will include fluctuations. For the saddle point we need to find and solve the classical equations of motion with the appropriate boundary terms. One notable feature of these equations, as pointed out in \cite{Tseytlin:2002ny}, is that for point-like solutions, i.e.\ with no $\sigma$ dependence, they are simply those of a particle moving in flat space. Thus, the complete set of geodesic solutions is the set of straight lines. 

For the light-cone gauge fixed theory, (corresponding to diagonal gauge),  the bulk equations of motion are given by 
\be
\label{eq:bulk_eom}
  &&
  \dot{x} = \frac{1}{s} p
  \comma
  \dot{\bar{x}} = \frac{1}{s} \bar{p}
  \comma
  \dot{\mathbf{z}} = \frac{1}{s} \mathbf{p}
  \comma
  \dot{x}^- = - \frac{1}{s} \Ham_{\mathrm{lc}}
  \; , \nn
\\ &&
  \dot{p} = - \frac{1}{s} \partial_\sigma \lrbrk{ \frac{\acute{x}}{\abs{\mathbf{z}}^4} }
  \comma
  \dot{\bar{p}} = - \frac{1}{s} \partial_\sigma \lrbrk{ \frac{\acute{\bar{x}}}{\abs{\mathbf{z}}^4} }
  \; , 
\\ &&
  \dot{\mathbf{p}} = - \frac{1}{s} \partial_\sigma \lrbrk{ \frac{\acute{\mathbf{z}}}{\abs{\mathbf{z}}^4} }
  - \frac{4}{s} \lrbrk{ \acute{x} \acute{\bar{x}} + \Half \acute{\mathbf{z}}^2 } \frac{{\mathbf{z}}}{\abs{\mathbf{z}}^6}
  \comma
  \dot{s} = 0
  \; . \nn
\ee
The boundary equations of motion at $\tau = \tau_1$ are 
\be
\label{eq:bd1_two_point}
  p = \frac{-1}{\sqrt{\lambda}} \frac{\delta \ln W_1}{\delta \bar{x}}
  \comma
  \bar{p} = \frac{-1}{\sqrt{\lambda}} \frac{\delta \ln W_1}{\delta x}
  \comma
  p_M = \frac{-1}{\sqrt{\lambda}} \frac{\delta \ln W_1}{\delta z_M}
  \comma
  s = \frac{-1}{\sqrt{\lambda}} \frac{\delta \ln W_1}{\delta x^-}
\ee
and those at $\tau = \tau_2$ are
\be
\label{eq:bd2_two_point}
  p = \frac{1}{\sqrt{\lambda}} \frac{\delta \ln W_2}{\delta \bar{x}}
  \comma
  \bar{p} = \frac{1}{\sqrt{\lambda}} \frac{\delta \ln W_2}{\delta x}
  \comma
  p_M = \frac{1}{\sqrt{\lambda}} \frac{\delta \ln W_2}{\delta z_M}
  \comma
  s = \frac{1}{\sqrt{\lambda}} \frac{\delta \ln W_2}{\delta x^-}
  \; ,
\ee
where
\be
   \frac{\delta \ln W}{\delta x} = - \Delta \, \frac{2(\bar{x}-\bar{a})}{\mathbf{z}^2+(\vec{x}-\vec{a})^2}~,~~
  \frac{\delta \ln W}{\delta \bar{x}} = - \Delta \, \frac{2(x-a)}{\mathbf{z}^2+(\vec{x}-\vec{a})^2} ~,\nn
  \ee
  and 
  \be
  \frac{\delta \ln W}{\delta x^-} =- \Delta \, \frac{2(\tau-a^+)}{\mathbf{z}^2+(\vec{x}-\vec{a})^2} ~\nn, 
  \ee
  and finally
  \be
  &  &\frac{\delta \ln W}{\delta z_M} = - \Delta \, \frac{\mathbf{z}^2 - (\vec{x}-\vec{a})^2}{\mathbf{z}^2 + (\vec{x}-\vec{a})^2} \frac{z_M}{\mathbf{z}^2} + J \, \frac{\mathbf{z}^2 \, n_M - (\mathbf{n}\cdot\mathbf{z})\, z_M}{\mathbf{z}^2 \, (\mathbf{n}\cdot\mathbf{z})} \; .
\ee
Let us consider the configuration where the two vertex operators are at the locations
\be
 \vec{a}_1 = (b_0,0,0,0) \comma \vec{a}_2 = (c_0,0,0,0)
\ee
with $c_0 > b_0$. This means that the vertex operators are separated only in the (Euclidean) time direction. More general configurations, equivalent up to boosts and rotations, can be treated at the cost of more complicated formulas. 
A solution to the equations of motion that does not have any $\sigma$-dependence is \cite{Tseytlin:2002ny}
\be \label{eqn:class-sol-2pt}
  x_\cl \eq 0
  \comma
  \bar{x}_\cl = 0
  \comma
  x^-_\cl = -\tau
  \comma
  s_\cl = \frac{\Delta}{\sqrt{\lambda}} \frac{i \sqrt{2}}{c_0 - b_0} \; , \\
  \mathbf{z}_\cl \eq \frac{1}{\sqrt{(\mathbf{n}_1-\mathbf{n}_2)^2}} \Big[ (c_0-x_0)e^{\phi}~ \mathbf{n}_1-(x_0-b_0)e^{-\phi} ~\mathbf{n}_2 \Big] \; . \nn
\ee
with $x_0 \equiv -i \sqrt{2} \tau$. The corresponding momenta follow from
\be
  p \eq s \dot{x}
  \comma
  \bar{p} = s \dot{\bar{x}}
  \comma
  \mathbf{p} = s \dot{\mathbf{x}} \; .
\ee
Thus we are required to set
\be \label{eqn:Virasoro-relation}
  J_1 = \Delta_1 = \Delta = \Delta_2 =  J_2 \; .
\ee
It is interesting to compare this to the corresponding computations in conformal gauge. In that case, the equations of motion do \emph{not} impose a relation between the dimension and the charge. There, this relation follows from demanding that the vertex operators are marginal operators. In our case, the origin of this relation is the Virasoro constraints. Since we have explicitly used them to eliminate the unphysical fields, we require that the vertex operators actually describe physical states satisfying the appropriate constraints. It is also worth mentioning that while we only demand that the solution satisfies the boundary conditions
at the worldsheet end-points, the fact is they are actually satisfied for any time. That is to say, the explicit
time dependence cancels in equations \eqref{eq:bd1_two_point} and \eqref{eq:bd2_two_point}.

Let us consider the two-point function for operators which carry the same $U(1)$ R-charge. That is we want to consider 
 ${\bf n}_1=-{\bf n}^\ast_2={\bf n}$ (as we will see taking the conjugate of $\mathbf{n}_2$ corresponds to this
 vertex being incoming), the solution \eqref{eqn:class-sol-2pt} then satisfies several useful relations:
\begin{align} \label{eqn:relations}
  \mathbf{z}_\cl^2 & = (x_0-b_0)(c_0-x_0) \; , &
  \mathbf{z}_\cl^2 + (\vec{x} - \vec{b})^2 & = (x_0-b_0)(c_0-b_0) \; , \\
  \dot{\mathbf{z}}_\cl^2 & = 2 \; , &
  \mathbf{z}_\cl^2 + (\vec{x} - \vec{c})^2 & = (c_0-x_0)(c_0-b_0) \; . \nn
\end{align}
as well as
\be 
  \mathbf{n}\cdot\mathbf{z}_\cl = (x_0-b_0) e^{-\phi}
  \comma
  \mathbf{n}^*\cdot\mathbf{z}_\cl = (c_0-x_0) e^{\phi}
  \; ,
\ee
and therefore
\be \label{eqn:recipocity}
  (\mathbf{n}\cdot\mathbf{z}_\cl) \, (\mathbf{n}^*\cdot\mathbf{z}_\cl) = \mathbf{z}_\cl^2
  \qquad\mbox{or}\qquad
  \frac{\mathbf{n}\cdot\mathbf{z}_\cl}{\abs{\mathbf{z}_\cl}} = \frac{\abs{\mathbf{z}_\cl}}{\mathbf{n}^*\cdot\mathbf{z}_\cl}
  \; .
\ee
Using this last relation in \eqref{eqn:def-genvertop} shows that changing the sign of $J$ is equivalent to complex conjugating $\mathbf{n}$. This is relevant for treating vertices of incoming and outgoing strings, which should correspond to taking complex conjugates. However, to put all vertices on the same footing we will for most part treat all vertices as outgoing but take the 
charges to be negative. 

Let us not impose the relations \eqref{eqn:Virasoro-relation} in the next few equations in order to see why they are important. The bulk action $\Action$ evaluates to zero even without these conditions while for the boundary action or, equally, the vertex operators we find that on the solution they contribute
\be
\label{eq:mod_vertex1}
  W_{1,\cl} \eq \frac{e^{-\phi J_1}}{(c_0 - b_0)^{{\Delta_1}}}  \, \lim_{x_0\to b_0}\left( \frac{x_0-b_0}{c_0-x_0}\right)^{\frac{\Delta_1-J_1}{2}} \; , \\
  \label{eq:mod_vertex2}
  W_{2,\cl} \eq  \frac{e^{-\phi J_2}}{(c_0 - b_0)^{{\Delta_2}}}  \,  \lim_{x_0\to c_0}\left( \frac{c_0-x_0}{x_0-b_0}\right)^{\frac{\Delta_2+J_2}{2}} \; .
\ee
%
Generically, these expressions would be infinity or zero because of the limit, however in the case $\Delta_1=J_1$ and $\Delta_2=-J_2$, corresponding to the protected BPS state, the vertex operators are
separately finite. More generally, when the  charges obey $\Delta_2=\Delta_1$ and $J_2=-J_1$ i.e. when the incoming and outgoing charges are the same, as they must be for the above solution, then we can write
\be
  \vev{ W_{1} W_{2} }_\cl 
   = \frac{{\cal N}}{\abs{c_0 - b_0}^{2\Delta}} \; .
\ee

In addition to the exponential components of the vertex operators there are the contributions from the $U(\tau)$ factors \eqref{eqn:BMN-vertex-op-U}. Writing these prefactors in the diagonal gauge $h^{ab}=\diag(\mathbf{z}^2,\mathbf{z}^{-2})$ and evaluating them on the above solutions for the times $\tau=\tau_{1,2}$, we find the simple result that $U_{\mathrm{I}}+U_{\mathrm{II}}+U_{\mathrm{III}}=1$. Thus, the two-point function to leading order in large $\sqrt{\lambda}$ is 
\be
\label{eq:two_pt_norm}
  \vev{ V_{1} V_{2} }_\cl 
   = \frac{{1}}{\abs{c_0 - b_0}^{2\Delta}} \; .
\ee

\section{Quantum two-point function}
\label{sec:quant-2pt}

We now wish to consider the effects of fluctuations about the saddle-point solutions considered above. This will allow us to find the corrections to the classical expressions for the vacuum and to consider the two-point functions of near-BMN states, that is operators with impurities. This requires including corrections to the vertex operators and including subleading corrections to the evaluation of the path integral. In doing so we follow methods standard from the analogous calculation of worldsheet correlation functions in flat space. This calculation is morally similar, and technically almost identical, to the quantization of fluctuations about  BMN strings and to the calculation of physical energies of such strings. In this case the action has Euclidean signature and the underlying classical solution has more parameters, however, as we shall see neither of these are significant. 
 
\paragraph{Fluctuation action.} We wish to determine the action for fluctuations of the coordinates and light-cone momentum parameter, $s$, where the expansion is in $\epsilon=\lambda^{-1/4}$, i.e. for a generic coordinate
\be
 X^\mu=X_{\rm cl}^\mu+\epsilon \tilde X^\mu~.
\ee
As the transverse momenta appear quadratically, imposing their equations of motion is equivalent, up to overall normalization constants, to performing the functional integration.
%
More specifically, the fluctuation expansion about the classical solution found in \secref{sec:class-2pt}.
is,
\footnote{For simplicity we focus on the solution corresponding to the two-point functions
with 
 ${\bf n}_1=-{\bf n}^\ast_2={\bf n}$. As we will see this is not a significant assumption as
 the fluctuation spectrum depends only on the overall charge $\Delta$ and not the boundary
 position or plane of rotation.}
\be
  x \eq \eps \tilde{x}
  \comma
  \bar{x} = \eps \tilde{\bar{x}} 
  \comma
  x^- = -\tau + \eps \tilde{x}^- 
  \comma
  s = s_\cl + \eps \tilde{s}
  \comma
  \mathbf{z} = \mathbf{z}_\cl + \eps \tilde{\mathbf{z}} 
  \; .
\ee
We plug this expansion \emph{only} into the bulk action and we will effectively treat the fluctuations at the boundary as if they vanished e.g. dropping total derivative terms. A rigorous treatment would also involve evaluating the Jacobian and functional derivatives involved in the coordinate redefinitions performed at intermediate steps, however as in flat space, these
should not be relevant to our considerations. As described in the previous section, the bulk action vanishes on the saddle point, so to zeroth order in $\eps$ the action will vanish. The first and second order terms in the expansion of the Lagrangian are
\be
  \Lagr \eq \eps \Bigsbrk{
    2 s_\cl \dot{\mathbf{z}}_\cl \cdot \dot{\tilde{\mathbf{z}}}
    + s_\cl \dot{\tilde{x}}^-
  } \nl
  + \eps^2 \lrsbrk{
      s_\cl \dot{\tilde{x}} \dot{\tilde{\bar{x}}}
    + \half s_\cl \dot{\tilde{\mathbf{z}}}^2
    + \tilde{s} \dot{\tilde{x}}^-
    + \tilde{s} \dot{\mathbf{z}}_\cl \cdot \dot{\tilde{\mathbf{z}}}
    + \frac{\acute{\tilde{x}}\acute{\tilde{\bar{x}}}+ \half \acute{\tilde{\mathbf{z}}}^2 }{s_\cl (x_0-b_0)^2(c_0-x_0)^2}
  } \; .
\ee
The order-$\eps$ terms are a total derivative because $s_\cl$ and $\dot{ \mathbf{z}}_\cl$ are constant. The order-$\eps^2$ can also be simplified: firstly, this is the only place where $\tilde{s}$ occurs and since it occurs linearly, we can integrate it out. Its equation of motion imposes the constraint on the zero-mode of ${\tilde x}^-$
\be
  \dot{\tilde{x}}^- = - \dot{\mathbf{z}}_\cl \cdot \dot{\tilde{\mathbf{z}}} \; .
\ee
The action for the quadratic 
fluctuations is thus
\be
  \Action_{\rm fl} = \frac{\sqrt{\lambda}}{2\pi} \int_{0}^{2\pi}\!d\sigma \int_{\tau_1}^{\tau_2}\!d\tau\:  \eps^2 \biggbrk{
      s_\cl \dot{\tilde{x}} \dot{\tilde{\bar{x}}}
    + \half s_\cl \dot{\tilde{\mathbf{z}}}^2
    + \frac{\acute{\tilde{x}}\acute{\tilde{\bar{x}}}+ \half \acute{\tilde{\mathbf{z}}}^2 }{s_\cl (x_0-b_0)^2(c_0-x_0)^2}
    }
\ee
Secondly, we redefine the fluctuations according to
\be
  \tilde{x} = \sqrt{F(\tau)} \, \tilde{\tilde{x}}
  \comma
  \tilde{\mathbf{z}} = \sqrt{F(\tau)} \, \tilde{\tilde{\mathbf{z}}}
\ee
with $F(\tau) = (x_0-b_0)(c_0-x_0) = \mathbf{z}_\cl^2$, then, dropping the tildes, we find
\be
  \Action_{\rm fl} \eq \frac{\sqrt{\lambda}}{2\pi} \int_{0}^{2\pi}\!d\sigma \int_{\tau_1}^{\tau_2}\!d\tau\: \eps^2 \biggbrk{
      s_\cl F \dot{x} \dot{\bar{x}}
    + \frac{1}{s_\cl F} \acute{x}\acute{\bar{x}}
    - \frac{s_\cl}{2F} (c_0-b_0)^2 x\bar{x} \nl \hspace{45mm}
    + \Half s_\cl F \dot{\mathbf{z}}^2
    + \frac{1}{2s_\cl F} \acute{\mathbf{z}}^2 
    - \frac{s_\cl}{4F} (c_0-b_0)^2 \mathbf{z}^2 
    } \; ,
\ee
where we have integrated by parts in $\tau$ and dropped the surface terms. We can redefine $\tau$ according to
\be
  \frac{d\tau}{s_\cl F(\tau)} = d\tilde{\tau}
  \comma
  s_\cl F(\tau) \partial_\tau = \partial_{\tilde{\tau}} \; ,
\ee
so that ($x_0 \equiv -i \sqrt{2} \tau$)
\be
\tilde \tau=\frac{1}{2}\left(\frac{\sqrt{\lambda}}{\Delta}\right)\ln \left(\frac{x_0-b_0}{c_0-x_0}\right)~~~,~ {\rm i.e.}~~~\left\{ 
\begin{array}{cc}
 \tilde \tau\rightarrow \infty &~{\rm as}~ x_0\rightarrow c_0 \\
 \tilde \tau\rightarrow -\infty &~{\rm as}~ x_0\rightarrow b_0
\end{array}\right.
\ee
then finally
\be \label{eqn:action-fl-simplified}
  \Action_{\rm fl} 
&=&\frac{1}{2\pi} \int d\sigma d{\tau}\:  \biggbrk{
     \Half \dot{X}{}^2
    + \Half \acute{X}^2 
    + \frac{1}{2}\mu^2 {X}{}^2
    } \; .
\ee
We have combined the transverse coordinates, $x, \bar x , \mathbf{z}$, into $X^I$, $I=1,\dots 8$, used the fact that $\epsilon^2=\lambda^{-1/2}$ and once again dropped the tildes, on this occasion from the time coordinate. Thus we find, as expected, 
the transverse massive scalars familiar from the BMN string where the mass of the fluctuations
is $\mu=\Delta/\sqrt{\lambda}$. While in general the dimension is a non-trivial function of
the coupling, $\Delta=\Delta(\sqrt{\lambda})$,  in the case at hand
\be
\Delta=\sqrt{\lambda} {\cal J}+ \order(1) \; ,
\ee
where ${\cal J}$ is the worldsheet density of the string angular momentum and so we can, to the order of interest, replace  $\mu={\cal J}$. One notable feature is that the fluctuation action only depends on the total charge of the classical string, 
$\Delta$, and not  on the positions of the vertex operators or on the specific plane in which the string is rotating on the S$^5$. It is possible to include the fermions and the fluctuation analysis is described in appendix \ref{app:fermions_fluc}, the result being that to quadratic order the action is again that of the BMN string with masses  $\pm \Delta/\sqrt{\lambda}$. We will for the most part focus on the bosonic calculation and only briefly mention the (non-trivial) extension to include fermions. 

\paragraph{Oscillator expansion.} We introduce an oscillator expansion of the fluctuation fields, 
\be
X^I = \sum_{n=-\infty}^{\infty} \frac{i}{\sqrt{2\omega_n}}(\alpha^I_n -\alpha^{I \dagger}_{-n}) e^{-in \sigma} \; ,
\ee
where $\omega_n=\sqrt{n^2+\mu^2}$. The corresponding canonical momenta are
\be
P^I = \frac{1}{2\pi} \sum_{n=-\infty}^{\infty}\sqrt{\frac{\omega_n}{2}}(\alpha^I_n +\alpha^{I\dagger}_{-n}) ~e^{-in\sigma}~,
\ee
and the bosonic light-cone Hamiltonian\footnote{When deriving the canonical momentum and the Hamiltonian from \protect\eqref{eqn:action-fl-simplified}, one has to be careful because this is the Euclidean action. It is probably easiest to temporarily Wick-rotate to Lorentzian signature.} is (up to a constant which cancels with the fermionic contribution)
\be
 H_{\rm lc}=\frac{1}{2\pi } \int d\sigma~ {\cal H}_{\rm lc}
=\sum_n \omega_n~ \alpha^{I\dagger}_n \alpha^{I}_n \; .
\ee

%
In defining the vertex operators we must now also include subleading terms characterizing the excitations at the boundary about the classical solution. That is, for every boundary labeled by $i=1,\dots,N$, we include a wave-function, $\psi_i( X_i^I)e^{ H_{{\rm lc},i} \tau_i} $, where $ X^I_i= X^I(\sigma, \tau_i)$ are the transverse boundary fluctuations and $\tau_i$ goes to minus infinity for incoming states and plus infinity for outgoing. It will be useful to expand the fluctuation momenta for each string boundary in terms of an oscillator basis, $\{a^I_{i,n\not=0},X^I_{i,0},P^I_{i,0}\}$, different than that used above\footnote{These oscillators still satisfy the usual commutation relations $[a^I_{i,n}, a^{J\dagger}_{j,m}]=\delta^{IJ}\delta_{nm}\delta_{ij}$.},
\begin{align}
a^I_{i,n} & = \frac{1}{\sqrt{2}} \lrbrk{ \alpha^I_{i,n}+\alpha^I_{i,-n} } \; , &
a^I_{i,-n} & = \frac{1}{i\sqrt{2}} \lrbrk{ \alpha^I_{i,n}-\alpha^I_{i,-n} } \; ,
\qquad (n=1,2,3,\ldots) \\
X^I_{i,0} & = \tfrac{i}{\sqrt{2\mu_i}}(\alpha^I_{i,0}-\alpha^{I\dagger}_{i,0}) \; , & 
P^I_{i,0} & = \sqrt{\tfrac{\mu_i}{2}}(\alpha^I_{i,0}+\alpha^{I\dagger}_{i,0}) \; ,
\end{align}
so that
\be
  X^I_i \eq X^I_{i,0}+\sqrt{2} \sum_{n=1}^\infty \lrbrk{ X^I_{i,n} \cos n\sigma+X^I_{i,-n}\sin n\sigma } \; , \nn \\
  P^I_i \eq \frac{1}{2\pi} \Bigsbrk{ P^I_{i,0}+\sqrt{2} \sum_{n=1}^{\infty} \lrbrk{ P^I_{i,n} \cos n\sigma+P_{i,-n}^I \sin n \sigma } }
\ee
with $X^I_{i,n}=\tfrac{i}{\sqrt{2\omega_{i,n}}}(a^I_{i,n}-a_{i,n}^{I\dagger})$ and  $P^I_{i,n}=\sqrt{\frac{\omega_{i,n}}{2}}(a^I_{i,n}+a_{i,n}^{I\dagger})$.

\paragraph{Worldsheet correlation functions.} Having shown that the fluctuations are described by the standard plane-wave action, calculating the quantum corrections to the two-point function is straightforward. To be slightly more general than necessary for a moment, as it will be useful later, we consider the $N$-point function. The string worldsheet corresponds to multiple segments which intersect at some specified times and locations, $\tau_r$ and $\sigma_r$, giving rise to $2N-4$ parameters describing the intersection points. In general we integrate over all such moduli however by using the invariance under global shifts of the coordinates we can fix the location of one intersection point. Thus for two-point and three-point functions, there will be no such integrations.  The $N$-point function is given by\footnote{This is not completely correct. As we shall see once we include fermions it is necessary to include additional factors at the string intersection points exactly analogous to the flat space case.}
\be
\label{eq:Npointfunct}
 \langle V_1(\tau_1) \kern-1pt& &\kern-14pt V_{2}(\tau_{2})\dots V_N(\tau_N)\rangle={\cal N}
 e^{-{\cal S}_{\rm cl}- {\cal B}_{{\rm cl}}}\int\prod^{N-3}_{r=1}d\tau_rd\sigma_r
 \int \prod {\cal D}{ X}^I \prod_{i=1}^N \psi_{i}( X_i^I)e^{{H_{{\rm lc},i}} \tau_i} e^{-\Action_{\rm fl}}~,\nn\\
 &=&{\cal N}e^{-{\cal S}_{\rm cl}-{\cal B}_{{\rm cl}}}
\int \prod_{i,n, I} dP^I_{i,n} ~\psi_{i}( P_{n,i}^I) 
\nn\\
& &\times
\int \prod {\cal D}{ X}^I  ~
{\rm exp}\Big[\sum_i H_{ {\rm lc},i}  |\tau_i|\Big]
~{\rm exp}\Big[{i\sum_i\int d\sigma~ P_i^I(\sigma)X^I(\sigma,\tau_i) -\Action_{\rm fl}}\Big]~,
 \ee
where in the last line we have Fourier transformed the wavefunctions to momentum space\footnote{One difference from the flat space version is that the integration over momenta includes the zero modes. For the pp-wave string the zero-modes are also harmonic oscillators and are on essentially the same footing as all other modes.}. ${\cal S}_{\rm cl}$ and ${\cal B}_{\rm cl}$ are the actions evaluated on the classical solutions. 
 
Now we can follow standard procedure from functional light-cone methods and integrate out the transverse coordinates. 
 \be
 \label{eq:npointNeu}
& &\langle V_1(\tau_1)V_{2}(\tau_{2})\dots V_N(\tau_N)\rangle={\cal N}e^{-{\cal S}_{\rm cl}- {\cal B}_{\rm cl}}\left[{\rm det}~\Delta\right]^{-4}\\
& & \qquad \qquad~~~\times \int \prod_{i,n,I} dP^I_{i,n} \prod_i \psi_{k_i}(P^I_{i,n})~~e^{\sum_i H_{{\rm lc},i} |\tau_i| +\frac{1}{4}\sum_{i,j}\int d\sigma'd\sigma''~ 
 P^I_i(\sigma')N(\sigma',\tau_i;\sigma'',\tau_j)P^I_j(\sigma'')}~.\nn
 \ee
For the wavefunction, we take, 
 \be
 \psi_{k_i}(P_{i,n}^I)=\prod_{n,I}\langle\Omega^{(i)}|(A^I_{i,n})^{k_{i,n}}|P^I_{i,n}\rangle ~.
 \ee
 where $|\Omega^{(i)}\rangle$ is the string vacuum state at each worldsheet boundary and $A^I_{i,n}$
 are the exact annihilation operators, and $ |P_{i,n}^I\rangle $ are momentum eigenstates.  To leading order in the $\sqrt{\lambda}$ expansion we simply
 have 
 \be
 |\Omega^{(i)}\rangle=|0\rangle^{(i)}+{\cal O}(\lambda^{-1/2})~,~~~{\rm and}~~~
 A^I_{i,n}=\alpha_{i,n}^I+{\cal O}(\lambda^{-1/2})~,
 \ee
 where $|0\rangle^{(i)}$ is the usual Fock vacuum for plane-wave oscillators, and $\alpha_{i,n}^I$
 are the corresponding oscillator annihilation operators (note these are the BMN oscillators not those in which we expanded the momenta). The momentum eigenstates are given by their usual expressions in terms of harmonic oscillators. 

\paragraph{Two-point function.} We now restrict ourselves to the two-point function where the worldsheet is simply a cylinder with 
the two states at the corresponding boundaries: state ``1" is incoming, $\tau_1\to-\infty$, and state ``2" outgoing, $\tau_2\to \infty$. The vertex operators depend on charges $\Delta_{1,2}$ and $J_{1,2}$ which are related, by demanding the physical state conditions, so that
\be
\Delta_{1,2}=|J_{1,2}|-P_{+;1,2}
\ee
where $P_+=-\tfrac{1}{P_-}H_{{\rm lc},i}$. As $H_{{\rm lc},i}$ are all of order unity they do not affect the classical saddle point about which we expand. Importantly, this implies that the classical worldsheet 
sourced by vertex operators for the BMN vacuum and those for near-BMN excited states are the same. Thus even for 
near-BMN strings we can use the analysis of the previous section 
and from the boundary action we have the classical contribution to the two-point function,
\footnote{We will not explicitly evaluate the functional determinant but simply absorb it into the 
overall normalization. It does not depend on the excitations of the string state and so 
has no dependence on the mode numbers.}
\be
\langle V_1(\tau_1)V_2(\tau_2)\rangle_{\rm cl}=\frac{{\cal N}}{|c_0-b_0|^{2\Delta}}~.
\ee
Turing to the quantum fluctuations, for generic points, solving for the worldsheet
Green's function is straightforward,
\be
N(\sigma,\tau;\sigma', \tau')&=&-\sum_{n=-\infty}^\infty \frac{2}{\omega_n}e^{-\omega_n|\tau-\tau'|}e^{i n(\sigma-\sigma')}~,\\
&=&-\frac{2}{\mu}e^{-\mu|\tau-\tau'|}-\sum_{n=1}^\infty \frac{4}{\omega_n}e^{-\omega_n|\tau-\tau'|}
(\cos n \sigma \cos n \sigma'+\sin n \sigma \sin n \sigma')~.\nn
\ee
Calculating the Green's function between points on the string endpoints one must include the effects of  waves reflected from the string boundary which in effect doubles those terms involving $e^{-\omega|\tau-\tau'|}$. 

We can thus rewrite the two-point function as 
\footnote{Note that the states are written in a shorthand and are strictly states in the tensor product of 
 Hilbert spaces. E.g. $|\{k_{1,n}\}, \{k_{2,n}\}\rangle=|(\alpha_{1,n}^I)^{k_{1,n}}\rangle^{(1)}\otimes| (\alpha_{2,n}^I)^{k_{2,n}}\rangle^{(2)}$.}
\be
\label{eq:quant_two_point}
\langle V_1(\tau_1)V_2(\tau_2)\rangle
=\frac{{\cal N}}{|c_0-b_0|^{2\Delta}}
   \langle  \{k_{2,n}\},\{k_{2,n}\}| 
e^{\sum_i H_{{\rm lc},i}\tau_i}e^{-2
 \sum_{n} \frac{1}{\omega_{n}}P_{1,n}^I P_{2,n}^Ie^{-\omega_n(\tau_2-\tau_1)}}|0\rangle~.
\ee
In the simplest case where both strings are in the vacuum state, i.e. $k_{i,n}=0$ for every $i$ and $n$, the 
light-cone Hamiltonian $P_{+,i}=0$ and, 
as $|\tau_1-\tau_2|=|\tau_1|+|\tau_2|\rightarrow (2\times \infty)$ with $\omega_n>0$ for each $n$, we simply find
\be
\langle V_1(\tau_1)V_2(\tau_2)\rangle
=\frac{1}{|c_0-b_0|^{2\Delta}}
\ee
where we have fixed the normalization ${\cal N}=1$. In the case where there are excitations, in \eqref{eq:quant_two_point}
we need to commute the light-cone Hamiltonian through the momentum operators. 
After doing this, the only  terms which are not exponentially suppressed are those of the
form $e^{-\sum_{n} a_{2,n}^{I\dagger} a_{1,n}^{I\dagger} }$, or switching to the BMN oscillators,  $e^{-\sum_{n} \alpha_{2,n}^{I\dagger} \alpha_{1,n}^{I\dagger} }$, so that 
\be
\langle V_1(\tau_1)V_2(\tau_2)\rangle
&=&\frac{1}{|c_0-b_0|^{2\Delta}}
   \langle \{k_{1,n}\},\{k_{2,n}\}| 
e^{-\sum_{n} \alpha_{2,n}^{I\dagger} \alpha_{1,n}^{I\dagger} }|0\rangle~\nn\\
&=&\frac{\delta_{\{k_{1,n},k_{2,n}\}}}{|c_0-b_0|^{2\Delta}}~.
\ee
\section{Classical three-point function}
\label{sec:class-3pt}
Our main concern is the generalization of previous the consideration to three point functions and it is
to this topic we now turn. We consider three vertex operators $V_{\Delta_1,J_1}(\tau_1,\vec{a}_1,\mathbf{n}_1)$, $V_{\Delta_2,J_2}(\tau_2,\vec{a}_2,\mathbf{n}_2)$, and $V_{\Delta_3,J_3}(\tau_3,\vec{a}_3,\mathbf{n}_3)$ corresponding
to three string states all with large charges sourcing a classical worldsheet. 
We will think of the string, with charges $\Delta_1, J_1$ as originating at the  boundary coordinate $\vec{a}_1$, extending into the bulk,  splitting at the bulk point $(\vec{x}_\nt,\mathbf{z}_\nt)$ into two parts with charges $\Delta_2, J_2$ and $\Delta_3,J_3$, and the two fragments reaching the boundary at locations $\vec{a}_2$ and $\vec{a}_3$, respectively. Thus there are three boundary actions and three string segments which reach from the three points on the boundary to the intersection point.
In addition each string segment is characterized by an internal coordinate ${\bf n}_i$ characterizing its motion
on the sphere. 

\begin{figure}%
\begin{center}
\includegraphics[scale=1]{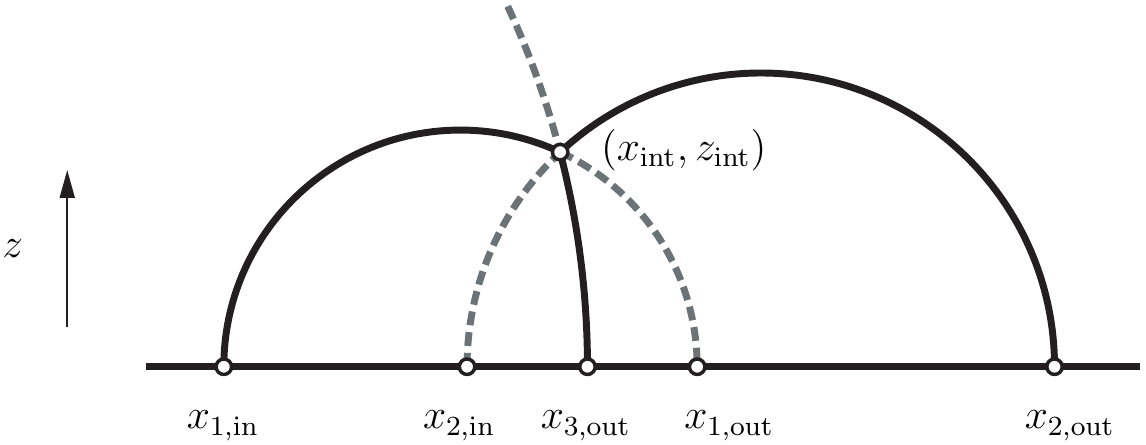}
\end{center}
\caption{\textbf{Three-string junction.} Each segment of the saddle-point solution for three strings is the saddle-point solution for two strings that we found in \protect\secref{sec:class-2pt}. However, from the two-string solution, we discard the part (dashed line) that lies beyond the intersection point.}%
\label{fig:strings1d}%
\end{figure}

We focus on the particular configuration where the vertex operators are aligned along the (Euclidean) time\footnote{We could write the time components as $b_0$, $c_0$, etc., but for convenience and clarity we drop 
the component index.}
\be
\label{eq:3pt_bd}
  \vec{a}_1 = (b,0,0,0) \comma
  \vec{a}_2 = (c,0,0,0) \comma
  \vec{a}_3 = (d,0,0,0) \; .
\ee
Since ``1'' is the in-string, and ``2'' and ``3'' are the out-strings,\footnote{We stress again that while we refer to the segments as ``incoming'' and ``outgoing'' we are working in a Euclidean formulation so the individual string segments are not physical propagating string solutions.} the intersection point will satisfy $b < x_\nt < c , d$ and we also choose $c < d$. The classical solution will look qualitatively like \figref{fig:strings1d}.

For each segment, we can recycle the solution \eqref{eqn:class-sol-2pt} that we found in the case of the two-point function
that is, for each segment $i=1,2,3$, the solution is of the form
\be
\label{eq:3ptansatz}
  x_{i,\cl} \eq 0
  \comma
  \bar{x}_{i,\cl} = 0
  \comma
  x^-_{i,\cl} = -\tau
  \comma
  s_{i,\cl} = \frac{\Delta_i}{\sqrt{\lambda}} \frac{i \sqrt{2}}{x_{i,\mathrm{out}} - x_{i,\mathrm{in}}} \; , \\
  \mathbf{z}_{i,\cl} \eq \frac{1}{|\mathbf{n}_{i,\rm in}-\mathbf{n}_{i,\rm out}|} \Big[ (x_{i,\mathrm{out}}-x)e^{\phi_i} \mathbf{n}_{i,{\rm in}}-(x-x_{i,\mathrm{in}})e^{-\phi_i} \mathbf{n}_{i,{\rm out}} \Big]  \; . \nn
\ee
where $x \equiv -i \sqrt{2} \tau$. If $x$ takes values along the entire interval $[x_{i,\mathrm{in}},x_{i,\mathrm{out}}]$, then this solution describes a semi-circle from the point $(x,z) = (x_{i,\mathrm{in}},0)$ to $(x_{i,\mathrm{out}},0)$. That was appropriate for the two-point function. For the three-point function only one end of each segment necessarily reaches all the way to the boundary while the other end will terminate at the intersection point which is generically in the bulk; therefore we need to restrict the interval for $x$ along each string segment, see \figref{fig:strings1d}. However, the solution describing each segment still depends on the this ``virtual" end-point which is a parameter determined by demanding that the strings intersect at the point $(\vec{x}_\nt,\mathbf{z}_\nt)$.
Similarly, for each segment one of $ \mathbf{n}_{i,{\rm in}}$ or  $\mathbf{n}_{i,{\rm out}}$ is determined
by the vertex operator on the boundary but the solution also depends on a ``virtual" vector which is again 
determined by demanding that the strings intersect. \footnote{In a previous version of this 
paper this freedom was neglected, resulting in an insufficiently general ansatz and hence an incorrect saddle point
for the sphere coordinates for non-extremal correlators. The correct ansatz, in conformal gauge, was found 
in \cite{Buchbinder:2011jr} and the correct saddle point was identified. We thank A. Tseytlin for bringing this to our attention.}

From the first property of the solution given in \eqref{eqn:relations} (which also holds for the more general ansatz 
\eqref{eq:3ptansatz}) we have the relation, satisfied along each segment, 
\be
  \mathbf{z}_\nt^2 & = (x_\nt-x_{i,\mathrm{in}})(x_{i,\mathrm{out}}-x_\nt) \; .
\ee
This allows us to eliminate the unphysical endpoint for string ``1'' and the initial points for strings ``2'' and ``3''. Thus we have:
\be
\label{eqn:table}
	\begin{tabular}{c|c|c|c}
Segment $i$ & Parameter $x$ & $x_{i,\mathrm{in}}$ & $x_{i,\mathrm{out}}$ \\ \hline
$1$ & $[b,x_\nt]$ & $b$ & $x_\nt + \frac{\mathbf{z}_\nt^2}{x_\nt-b}$\\
$2$ & $[x_\nt,c]$ & $x_\nt - \frac{\mathbf{z}_\nt^2}{c-x_\nt}$ & $c$ \\
$3$ & $[x_\nt,d]$ & $x_\nt - \frac{\mathbf{z}_\nt^2}{d-x_\nt}$ & $d$
  \end{tabular}
\ee
By construction we have ensured that the segments meet in the AdS$_5$ subspace; all segments have the point $(x_\nt, z_\nt)$ in common, where $z_\nt = \abs{\mathbf{z}_\nt}$. For the segments to meet on the sphere, we need to further impose
\be \label{eqn:intersection}
\mathbf{z}_{\rm int}= \left. \mathbf{z}_{1,\cl}\right|_{x=x_{\rm int}} =\left. \mathbf{z}_{2,\cl}\right|_{x=x_{\rm int}}= \left. \mathbf{z}_{3,\cl}\right|_{x=x_{\rm int}} \; .
\ee
Using this condition we can determine, for example, on the first segment
\be
{\bf n}_{1,\rm out}=\frac{e^{2\phi_1}}{(x_{\rm int}-b)^2}\Big[ z_{\rm int}^2 {\bf n}_{1,\rm in} - 2({\bf n}_{1,\rm in}\cdot{\bf z}_{\rm int}){\bf z}_{\rm int}\Big]
\ee
with similar expressions for ${\bf n}_{2,\rm in}$ and ${\bf n}_{3,\rm in}$.

For the above solution we require that $\Delta_1 = J_1$, $\Delta_2 =J_2$, and $\Delta_3 = J_3$ so that, plugging the solution
 into the action, one finds, as for the two-point funciton,  that the bulk action vanishes, the prefactors become unity and the only non-trivial  contribution comes from the boundary terms which, after some algebra, give
\be
\label{eqn:bdyaction3pt} 
  \mathcal{B} \eq \mathcal{B}_{\rm AdS}+\mathcal{B}_{\rm Sph}
  \ee
  where, with the notation $a_{0,1}=b$, $a_{0,2}=c$ and $a_{0,3}=d$,
  \be
\mathcal{B}_{\rm AdS}&=&  \sum_{i=1}^3\Delta_i \ln\frac{ |\mathbf{z}_\nt|}{\mathbf{z}_\nt^2 + (x_\nt-a_{0,i})^2}~,~~~
 \mathcal{B}_{\rm Sph}=\sum_{i=1}^3 \Delta_i \ln\frac{\mathbf{n}_i \cdot \mathbf{z}_\nt}{ |\mathbf{z}_\nt|} \; .
\ee
It looks like we have essentially gone back to \eqref{eqn:def-genvertop}. However, there is an important difference: the $\vec{x}(\tau)$ and $\mathbf{z}(\tau)$ in \eqref{eqn:def-genvertop} are to be evaluated at the boundary. In \eqref{eqn:bdyaction3pt} the point $(x_\nt,\mathbf{z}_\nt)$ lies at the intersection of the three strings, a point generically in the bulk.
The remaining step in performing the semiclassical evaluation of the path-integral is to evaluate the saddle
point of the finite-dimensional integral over the undetermined intersection point, 
\footnote{Here, as we have 
 specified to the case where the vertex operators lie along a 
single boundary direction, $x_{\rm int}$,  the integration is over an AdS$_2 \subset$AdS$_5$. 
It is straightforward to generalize to the full AdS$_5$ space.}
\be 
  \langle V_1(\tau_1) V_2(\tau_1) V_3(\tau_1)\rangle 
  &=&\int dx_{\rm int} dz_{\rm int}~e^{-\mathcal{B}_{\rm AdS}}
  \int d^5\Omega_{\rm int}~\prod_{i=1}^3 \left(\frac{{\bf n}_i \cdot {\bf z }_{\rm int} }{|{\bf z}_{\rm int }|}\right)^{\Delta_i}~.
\ee
By making use of the standard parametrisation of the five sphere
\be
\label{eq:sphere_param}
\frac{z_1+i z_2}{|\bf{z}|}=\cos \gamma e^{i\beta_1}~,~~\frac{z_3+i z_4}{|\bf{z}|}=\sin \gamma \sin \psi e^{i\beta_2}~,~
~\frac{z_5+i z_6}{|\bf{z}|}=\sin \gamma \cos \psi e^{i\beta_3}
\ee
we factorize  the integration into an AdS part and a sphere part. 
We perform the saddle-point evaluation of the AdS integral by, as discussed in
\eqref{eqn:intersection-point}, extremizing the function ${\mathcal B}_{\rm AdS}$. The integral 
over the sphere depends on the choice of vectors ${\bf n}_i$.
The degenerate case where the three vertex operators correspond to strings rotating in the same plane is dual to the extremal three-point functions in the gauge theory. More generally the strings can rotate in orthogonal planes or in diagonal combinations.

\paragraph{Extremal correlator.} Let 
\be
 \mathbf{n}_1 = \mathbf{n}^\ast_2 = \mathbf{n}^\ast_3 = (1,i,0,0,0,0) \; .
\ee
In this case only the first two components of ${\bf z}_{\rm int}$ appear in the minimization problem and we can take  $\mathbf{z}_\nt = (z_{\nt,1},z_{\nt,2},0,0,0,0)$. Let the intersection coordinates be $z_{\nt,1} = z_\nt \cosh \varphi_\nt$ and $z_{\nt,2} = z_\nt \sinh \varphi_\nt$. Then $\mathbf{n}_1 \cdot \mathbf{z}_\nt = z_\nt e^{\varphi_\nt}$, $\mathbf{n}_2 \cdot \mathbf{z}_\nt = \mathbf{n}_3 \cdot \mathbf{z}_\nt = z_\nt e^{-\varphi_\nt}$ and the boundary action becomes
\be
  \mathcal{B} \eq  \Delta_1 \ln\frac{z_\nt}{\mathbf{z}_\nt^2 + (x_\nt-b)^2}
                  +\Delta_2 \ln\frac{z_\nt}{\mathbf{z}_\nt^2 + (x_\nt-c)^2}
                  +\Delta_3 \ln\frac{z_\nt}{\mathbf{z}_\nt^2 + (x_\nt-d)^2} \nl
                  +(\Delta_1 - \Delta_2 - \Delta_3) \varphi_\nt \; .
\ee
Here we see that the action depends on the direction $\varphi_{\rm int}$ linearly, as the strings can intersect anywhere along 
a circle. Hence there is no minima and performing the integration over the intersection point produces a delta function for the angular momenta imposing $J_1=J_2+J_3$, or for the solution we consider imposes  the constraint
\be \label{eqn:Delta-conservation}
  \Delta_1 = \Delta_2 + \Delta_3
\ee
thus these are extremal correlators. Minimizing for $x_\nt$ (or really $\vec{x}_\nt$ but there is only one non-trivial component) and $z_\nt$ yields the result presented in \eqref{eqn:intersection-point}. As previously mentioned, on this solution the bulk action vanishes, the prefactors become constants and the only contribution comes from the boundary terms yielding
\be
\label{eq:extremal3pt}
  \langle V_1(\tau_1) V_2(\tau_1) V_3(\tau_1)\rangle &=&
\frac{{\mathcal C}_{\rm AdS}}{\abs{b - c}^{\alpha_3} \abs{b - d}^{\alpha_2} \abs{c - d}^{\alpha_1} }
\ee
with 
\be
   \label{eqn:non-extremal}  
 {\mathcal C}_{\rm AdS} 
  \eq \left( 
  \frac{
      \alpha_1^{\alpha_1} \alpha_2^{\alpha_2} \alpha_3^{\alpha_3}
      (\alpha_1 + \alpha_2 + \alpha_3)^{\alpha_1 + \alpha_2 + \alpha_3}
}{
      (\alpha_1 + \alpha_2)^{\alpha_1 + \alpha_2}
      (\alpha_2 + \alpha_3)^{\alpha_2 + \alpha_3}
      (\alpha_3 + \alpha_1)^{\alpha_3 + \alpha_1}
} 
~\right)^{1/2}\nn\\
\ee
and the $\alpha$'s as in \eqref{eq:alphadef}. This calculation is done for generic $\Delta_i$'s. We now impose the constraint \eqref{eqn:Delta-conservation} for the extremal case and the above the result simplifies significantly
\be
  \langle V_1(\tau_1) V_2(\tau_1) V_3(\tau_1)\rangle \eq \frac{1}{\abs{b - c}^{\alpha_3} \abs{b - d}^{\alpha_2}} \; .
\ee

\paragraph{Non-extremal correlators.} The above methods are general enough to allow for the different strings to be rotating in different intersecting planes of the S$^5$. 
The AdS boundary action, $\mathcal{B}_{\rm AdS}$, remains the same and so the extremization is unchanged.
Minimization with respect to $z_{\rm int}$ and $x_{\rm int} $ yields the same result as before and thus the space-time dependence is equivalent to \eqref{eqn:non-extremal}, however, now the integration over the sphere is non-trivial
 \footnote{For large charges this integration can also be done by saddle-point approximation. This was performed
 in \cite{Buchbinder:2011jr} and the explicit result was found. Here, we leave the integration unperformed}.
 Generically, the result is 
 \be 
 \label{eq:3ptnonextremal}
  \langle V_1(\tau_1) V_2(\tau_1) V_3(\tau_1)\rangle &=&\frac{{\mathcal C}_{\rm AdS}}{\abs{b - c}^{\alpha_3} \abs{b - d}^{\alpha_2} \abs{c - d}^{\alpha_1} } 
\int d^5\Omega_{\rm int}~\prod_{i=1}^3 \left(\frac{{\bf n}_i\cdot {\bf z }_{\rm int} }{|{\bf z}_{\rm int }|}\right)^{\Delta_i} \; .
\ee
In general, we can expect to find non-vanishing non-extremal correlators
when 
\be
{\bf n}_1\cdot {\bf n}_2\neq 0~, ~{\bf n}_1\cdot {\bf n}_3\neq 0~, ~{\bf n}_2\cdot {\bf n}_3\neq0~.
\ee
This result is reminiscent to the harmonic superspace description of the three-point vacuum correlators in 
\cite{Beisert:2002tn}. In that case the space-time super-coordinates are augmented by an auxiliary bosonic coordinate
$V^m$, $m=1,\dots, 6$ such that $V^2=0$, $ V\cdot V^\ast=1$ c.f. \eqref{eqn:properties-n}. In terms of the scalar fields of ${\cal N}=4$ SYM,
$\Phi_m$, the vacua are, schematically ${\cal O}_J={\rm Tr}(Z^J)$ and thus labelled by a choice of $V$, $Z=\Phi_mV^m$. The three-point 
function of three different vacua is given by
\be
\langle {\cal O}_{J_1}{\cal O}_{J_2}{\cal O}_{J_3}\rangle= C^{123} K_{12}^{J_1+J_2-J_3/2}K_{23}^{J_2+J_3-J_1/2}K_{13}^{J_1+J_3-J_2/2}
\ee
where the leading bosonic component is
\be
 K_{12}\sim \frac{\delta_{mn}V_1^m V_2^n}{(x_1-x_2)^2}~.
 \ee

\paragraph{Comparison with weak coupling.} It is also interesting to compare the normalization of the non-extremal result  \eqref{eqn:non-extremal} 
 with that computed at weak coupling in gauge theory and at strong coupling using  the supergravity approximation \cite{Lee:1998bxa}. For three chiral primary operators, $O^{I_i}$, with dimensions $\Delta_{i}=J_i$, defined such that,
  \be
  \langle O^{I_1}(\vec{a}_1)O^{I_2}(\vec{a}_2)\rangle=\frac{\delta^{I_1I_2}}{|\vec{a}_1-\vec{a}_2|^{2\Delta_1}}
  \ee
that is, with the normalization as in \eqref{eq:two_pt_norm}, the three-point function is given in the planar limit by 
\be
    \langle O^{I_1}(b)O^{I_2}(c)O^{I_3}(d)\rangle&=&
    \frac{1}{N}
    \frac{\sqrt{J_1 J_2 J_3}}{|b-c|^{\alpha_3}  |b-d|^{\alpha_2} |c-d|^{\alpha_1}}
    \biggsbrk{ \bigbrk{\frac{\alpha_1+\alpha_2+\alpha_3}{2}+2}! \,
   \frac{\tfrac{\alpha_1}{2}!\tfrac{\alpha_2}{2}!\tfrac{\alpha_3}{2}!}{J_1!J_2! J_3!} } \nn\\
 & &\times  \frac{ 1 }{2\pi^3}
   \int_{S^5} Y^{I_1}Y^{I_2}Y^{I_3}d\Omega
\ee
where the $Y^I$'s are the ultra-spherical harmonics normalized such that\footnote{This is a different normalization than that of \cite{Lee:1998bxa}.}
\be
\label{eq:harmonics_norm}
  \frac{1}{\pi^3}\int_{S^5} Y^{I_1}Y^{I_2}d\Omega=\frac{\delta^{I_1I_2}}{2(J_1+1)(J_1+2)}
\ee
and where $N$ is the rank of the gauge group. This expression is obviously different than  \eqref{eq:3ptnonextremal} even if we make appropriate choices for the ultra-spherical harmonics characterizing the operators. For example, in the extremal limit there is a numerator factor, $\sqrt{J_1J_2J_3}$, absent from the string calculation. However, the string calculation  assumes that all the charges, $J_i$, are large, i.e. $J_i= \sqrt{\lambda}{\cal J}_i$ with ${\cal J}_i={\cal O}(1)$. Moreover, we take the $\alpha_i$'s to be large which is natural from the string theory, as generically $\alpha_1=J_2+J_3-J_1\sim \sqrt{\lambda}$. We can now take the extremal limit $\alpha_1 \to 0$ or ${\cal J}_2+{\cal J}_3 -{\cal J}_1\to 0$ however we should be aware that it is after having already taken the large charge limit. Using Stirling's formula we can approximate the factorials $n!\sim n^n e^{-n}$ so that
\be
    \langle O^{I_1}(b)O^{I_2}(c)O^{I_3}(d)\rangle \eq
    \frac{g_s}{|b-c|^{\alpha_3}  |b-d|^{\alpha_2} |c-d|^{\alpha_1}}
    \biggsbrk{ \frac{(\alpha_1+\alpha_2+\alpha_3)^{(\alpha_1+\alpha_2+\alpha_3)}
\alpha_1^{\alpha_1}\alpha_2^{\alpha_2}\alpha_3^{\alpha_3}}{J_1^{J_1}J_2^{J_2} J_3^{J_3}} }^{1/2}\nn\\
& &\times\frac{1  }{\pi^3}
   \int_{S^5} Y^{I_1}Y^{I_2}Y^{I_3}d\Omega
\ee
which indeed reproduces \eqref{eq:3ptnonextremal} for specific choices of the three ultra-spherical harmonics
\footnote{In an earlier version of this paper, the dependence on the sphere was incorrect due to 
an insufficiently general ansatz. The correct treatment, in conformal gauge, was found \cite{Buchbinder:2011jr}.}
 up to the overall factor of the string coupling $g_s=\frac{1}{N}$ which we have omitted. In this approximation we have dropped all polynomial terms of
order ${\cal O}(J_i^n)$ for finite $n$ or rather we have set them equal to one. To find the correct prefactor 
agreeing  with \cite{Lee:1998bxa} it would most likely be necessary to include the full fermionic terms in the definition
of the $U(\tau_i)$ appearing in the definition of the vertex operator however to the order of our considerations it 
does not seem to contribute. 
\footnote{This possibility of this necessity was stressed to us by A. Tseytlin.} 

With regard to the integral over spherical harmonics, we have not considered the vertex operators for such general configurations from the string point of view, however in \appref{app:toy_non_ext} we briefly describe the point particle on AdS space as a toy model for BMN strings and show how the semiclassical three-point function is  indeed proportional to the overlap of ultra-spherical harmonics.
As a simple example we can consider the  extremal case,
\be
Y^{I_1}=\left(\frac{z_1-i z_2}{|\bf{z}|}\right)^{J_1}~, ~~Y^{I_{2,3}}=\left(\frac{z_1+i z_2}{|\bf{z}|}\right)^{J_{2,3}}
\ee
so that the overlap of three harmonics is just that of two harmonics both with $J=J_2+J_3$ and one can use the two-point formula \eqref{eq:harmonics_norm} which has the required behavior to match \eqref{eq:extremal3pt}.

\section{Circular winding strings}

Here we wish to repeat the three-point analysis for the circular winding string. Such string solutions were first considered in \cite{Frolov:2003qc} and the general class of solutions was described in \cite{Arutyunov:2003za}. In general, one can consider rigid string solutions with angular momenta in both the AdS$_5$, $S_r$, and S$^5$, $J_i$, spaces and with various windings $k_r$, $m_i$ in both subspaces. The simplest case is a particle in the AdS space with two equal angular momenta $J$  and winding $m$ on the sphere. These strings have a particularly simple relation between their energies and charges: $E=\sqrt{4 J^2+\lambda m^2}$. While this solution is unstable once quantum fluctuations are considered, it acts as an interesting probe of the integrable structures underlying the planar limit of the AdS/CFT duality.

In \cite{Frolov:2003qc} and  \cite{Arutyunov:2003za},  conformal gauge and global coordinates were used to 
describe the solution. In conformal gauge, but now using Poincar\'e coordinates,  the simplest circular winding solution (of the Lorentzian theory) is given by, using the notation $\mathbf{z}=(z_1,\ldots,z_6)$,
\be
z_1=\frac{1}{\sqrt{2}  \cos \kappa \tau}\cos(\omega \tau+m\sigma+\phi_1)~,& &~~~
z_2=\frac{1}{\sqrt{2}  \cos \kappa \tau}\sin(\omega \tau+m\sigma+\phi_1)\nn\\
z_3=\frac{1}{\sqrt{2}  \cos \kappa \tau}\cos(\omega \tau-m\sigma+\phi_2)~,& & ~~~
z_4=\frac{1}{\sqrt{2}  \cos \kappa \tau}\sin(\omega \tau-m\sigma+\phi_2)~,\nn\\
x_0\kern-5pt&= &\kern-5pt \tan \kappa\tau~, 
\ee
and $z_5=z_6=0$. 
This solution has equal angular momentum in the two orthogonal planes, 1-2 and 3-4, and winding $m$. The Virasoro constraints, satisfied by a physical solution, imply that $\kappa=\sqrt{m^2+\omega^2}$. This solution has energy and angular momentum
\be
E&=&\frac{\sqrt{\lambda}}{2\pi}\int d\sigma~ \frac{1}{z^2} \dot x_0=\sqrt{\lambda}\,\kappa \nn\\
J_{12}&=&\frac{\sqrt{\lambda}}{2\pi}\int d\sigma~ \frac{1}{z^2} \Big[z_2\dot z_1-z_1\dot z_2\Big]=-\sqrt{\lambda}\,\frac{\omega}{2}=J_{34}~.
\ee

\paragraph{Euclidean Solution.} One can straightforwardly check that the analytical continuation
is a solution of the Euclidean equations of motion.
The corresponding solution in diagonal gauge is, with $\kappa^2 = \omega^2 + m^2$,
\be
\label{eq:circ_sol_diag}
z_1=\frac{x_{\rm out}-x_{\rm in}}{2\sqrt{2} \cosh \kappa \tilde \tau} \cosh(\omega \tilde \tau+i m\sigma+\phi_1)~, & &~~~
z_2=i \frac{x_{\rm out}-x_{\rm in}}{2\sqrt{2} \cosh \kappa \tilde \tau} \sinh(\omega \tilde \tau+i m\sigma+\phi_1)~,\nn\\
z_3=\frac{x_{\rm out}-x_{\rm in}}{2\sqrt{2} \cosh \kappa \tilde \tau} \cosh(\omega \tilde \tau-i m\sigma+\phi_2)~, & &~~~
z_4=i \frac{x_{\rm out}-x_{\rm in}}{2\sqrt{2} \cosh \kappa \tilde \tau} \sinh(\omega \tilde \tau-i m\sigma+\phi_2)~,\nn\\
x^+=\tau~, ~~~~ x^-=-\tau ~, & &~~~ s = \frac{i\sqrt{2}\,\kappa}{x_{\rm out}-x_{\rm in}}
\ee
with 
\be
\label{eq:tautilde}
\tilde \tau =\frac{1}{\kappa}\arctanh \biggsbrk{ \frac{2x_0 - x_{\rm in} - x_{\rm out}}{x_{\rm out}-x_{\rm in}} } \comma x_0 \equiv -i \sqrt{2}\tau \; .
\ee
Setting $x_{\rm out}=1$ and $x_{\rm in}=-1$ this is exactly the analytic continuation of the global solution written in Poincar\'e coordinates but we have made the generalization to more arbitrary boundary conditions along the $x_0$ coordinate (even more
general solutions can be found by arbitrary boosts). 

This solution has the correct boundary behavior to correspond to the integrated versions of the vertex operators\footnote{As for the BMN string there is an issue regarding how we define the integrated vertex e.g.\ $\exp\big[\frac{1}{2\pi}\int d\sigma \ln V_i\big]$ but again, on the solution, the vertex does not depend on $\sigma$ so both definitions give the same answer.}
\be
\label{eq:simp_circ_vert}
  V^{\rm R}_{\Delta,J,J,m}(\vec{\sigma},\vec{a}) = \biggbrk{\frac{ \abs{\mathbf{z}} }{ \mathbf{z}^2 + (\vec{x} - \vec{a})^2}}^\Delta
                               \biggbrk{\frac{z_1+i z_2 }{ \abs{\mathbf{z}} }}^J  \biggbrk{\frac{z_3+i z_4 }{ \abs{\mathbf{z}} }}^J
 \ee
where, for physical states, we require $\Delta=\sqrt{4 J^2+\lambda m^2}$. The specific solution above corresponds to the vertex operators being located at positions $\vec{a}_1=(x_{\rm in},0,0,0)$ and $\vec{a}_2=(x_{\rm out},0,0,0)$ on the four-dimensional boundary. 
 
This vertex operator is essentially that of \cite{Ryang:2010bn} however it is not clear that this is in fact the correct form for the circular winding string\footnote{We are grateful to A. Tseytlin for very useful discussions on this and points related.}. Even if  we assume that the analytically continued solution \eqref{eq:circ_sol_diag} is the correct saddle-point for the two-point function, this does not uniquely determine the form of the vertex operator and there could be other vertex operators which produce the same boundary action. For example, if there are non-trivial polynomial terms, 
 \be
  V_{\Delta,J,J,m}(\vec{\sigma},\vec{a}) = \biggbrk{\frac{ \abs{\mathbf{z}} }{ \mathbf{z}^2 + (\vec{x} - \vec{a})^2}}^\Delta
                               \biggbrk{\frac{z_1+i z_2 }{ \abs{\mathbf{z}} }}^J  \biggbrk{\frac{z_3+i z_4 }{ \abs{\mathbf{z}} }}^J U_m(\vec{\sigma}) 
 \ee 
 with, schematically, 
 \be
 U_m(\vec{\sigma}) \sim \biggsbrk{{\bf z}^2h^{ab}\partial_a \left(\frac{z_1+i z_2}{|{\bf z}|}\right) \partial_b  \left(\frac{z_1-i z_2}{|{\bf z}|}\right) }^m \biggsbrk{{\bf z}^2h^{ab}\partial_a \left(\frac{z_3+i z_4}{|{\bf z}|}\right) \partial_b  \left(\frac{z_3-i z_4}{|{\bf z}|}\right) }^m~,
\ee
the boundary action is identical and so the same classical surface will be a solution, however the normalization of the two-point function is different. It is not clear that even this is sufficiently general. For example another proposal, one which is perhaps better motivated, was given in \cite{Buchbinder:2010gg}. It involves T-dualized coordinates dual to the angular coordinates along which the string is extended, $\beta_1$ and $\beta_2$, see \eqref{eq:sphere_param}. That is, we introduce variables $\tilde \beta_1$ and $\tilde \beta_2$ which are related by worldsheet
duality to the angles $\beta_1$ and $\beta_2$, 
\be
\partial_\sigma  \tilde \beta_{1,2}=-{\bf z}^2 \partial_\tau \beta_{1,2}~~~\text{and}~~~{\bf z}^2 \partial_\tau  \tilde \beta_{1,2}= \partial_\sigma \beta_{1,2} \; ,
\ee
where here we have dropped the dependence on the remaining coordinates as they will not be relevant. In terms of these variables the proposed vertex is
\be
V_{\Delta,J,J,m}(\vec{\sigma},\vec{a}) = \biggbrk{\frac{ \abs{\mathbf{z}} }{ \mathbf{z}^2 + (\vec{x} - \vec{a})^2}}^\Delta
                              e^{i J \beta_1}  e^{i J \beta_2}  e^{i \sqrt{\lambda}\frac{m}{2} \tilde \beta_1} e^{-i\sqrt{\lambda} \frac{m}{2} \tilde \beta_2} \; ,
\ee
which now has modified exponentially large terms. That this vertex provides a source for the circular winding string solution was shown for the theory defined on the plane in  \cite{Buchbinder:2010gg}. This proposal thus suffers from the same ambiguities as that of \cite{Ryang:2010bn}, for example there could be missing polynomial terms. However, given its explicit dependence on the winding parameter, and the interpretation of exchanging momentum for winding by T-duality,  it seems perhaps more likely correct.  For our purposes it is useful to consider the fields  at the boundary at time $\tau_i$
\be
 \left. \tilde \beta \right|_{\tau=\tau_i}&=&\int_{\sigma_0}^\sigma d\sigma' \partial_\sigma \tilde \beta (\sigma',\tau_i)=-\int_{\sigma_0}^\sigma d\sigma' z^2\partial_\tau  \beta(\sigma',\tau_i)\nn\\
 &\sim&\int_{\sigma_0}^\sigma d\sigma' ~p_\beta(\sigma',\tau_i) \; ,
\ee
where $\sigma_0$ is some reference point. Thus we interpret the vertex in our first order formalism as 
\be
V^{\rm B}_{\Delta,J,J,m}(\vec{\sigma},\vec{a}) = \biggbrk{\frac{ \abs{\mathbf{z}} }{ \mathbf{z}^2 + (\vec{x} - \vec{a})^2}}^\Delta
                              e^{i J \beta_1}  e^{i J \beta_2}  
                              e^{i \sqrt{\lambda}\frac{m}{2} \int_{\sigma_0}^\sigma d\sigma' ~ \Big[p_{\beta_2}(\sigma',\tau_i)-p_{\beta_1}(\sigma',\tau_i) \Big]} \; .
\ee
It is interesting to note that evaluated on the solution the angular momenta are equal, $p_{\beta_1}=p_{\beta_2}$,  which implies that  portion of the vertex depending on the T-dual coordinates becomes 
trivial. Moreover
it implies that, again on the solution, this vertex is $\sigma$-independent and so we can trivially interpret it as a contribution to the boundary action as in \eqref{eq:boundary_act}. We can now check the equations of motion including the boundary terms, whereupon we find that it does not satisfy the correct boundary conditions. This is essentially immediate from the fact that the 
vertex contributes to the $p_\beta$ equations of motion boundary terms which cannot be canceled by corresponding 
terms from the bulk action, see \eqref{eq:bulk_eom}. This is due to the choice of boundary conditions
for the bulk action \eqref{eqn:bulk-action}, in particular there is a choice in writing 
\be
{\cal S}=\frac{\sqrt{\lambda}}{2\pi}\int d\sigma d\tau~ \Big[p_A \dot x^A -{\cal H}\Big]\neq \frac{\sqrt{\lambda}}{2\pi}\int d\sigma d\tau~ \Big[\frac{p_A \dot x^A-\dot p_A x^A}{2} -{\cal H}\Big]~.
\ee
If we make the second choice for the coordinates $\beta_{1}$ and $\beta_{2}$, we now get additional boundary terms in the 
$p$ equations of motion and we can satisfy the boundary conditions
with the appropriate choice of $\sigma_0$'s. Given the formal nature of the T-duality and its global consequences it
is perhaps unsurprising that we must modify the boundary conditions of our string, however in doing so we also modify the relation between the vertex operator charges and the parameters of our solution. In particular, with the symmetric choice for the bulk action
$\tfrac{1}{2}(p_\beta\dot \beta-\dot p_\beta \beta)$ we find that $\omega=4J$. This is a different relation than for the analogous 
parameter in the physical solution and so also different than that found in  \cite{Buchbinder:2010gg}. In  \cite{Buchbinder:2010gg} a Lagrangian
approach using both the original fields and the dual fields at the same time was used. It is possible that 
in our first-order formalism calculation we too should use some doubled formalism, however we will leave this to another occasion.

The ambiguity in the choice for the vertex operator is related, at least in part, to the fact that the global charges are not sufficient to uniquely identify which state a given string solution corresponds to. If we had a better understanding of the higher, integrable charges of vertex operators we may be able to match them one-to-one to classical solutions. However, we currently do not have such a description and so we evaluate the correlators of three vertices of the type \eqref{eq:simp_circ_vert}. If there are polynomial terms to be added they would have to be evaluated on the classical solution while if the vertex is completely different even the boundary action contributions to the correlator may be different. 
 
There is another, related point: just as we have not proven that there is a unique vertex operator consistent with a given  solution, a given solution for a vertex operator may not be the global minimum of the action. There can be multiple contributions from different local minima and until one has a complete classification of solutions the saddle-point computation may be incomplete. In certain cases, when strings are BPS or when one can take a flat space limit it is possible to gain intuition regarding which 
solution dominates the path integral however the circular winding string does not have a smooth limit to either of these configurations. 

\paragraph{Three-point function.} Exactly parallel to the BMN case, we now consider a worldsheet consisting of three segments, each ending on a vertex operator, $V^{\rm R}(\tau_i, \vec{a}_i)$, at the boundary with charges $\Delta_i$, $J_i$ and $m_i$ such that $\Delta_i= (4 J_i^2+\lambda m_i^2)^{1/2}$. Our considerations would be identical if we allowed a polynomial prefactor $U(\tau_i)$ or if we used the vertices $V^{\rm B}(\tau_i, \vec{a}_i)$ but with alternative definition of the boundary conditions for the bulk action. For each segment $i=1,2,3$, we use the general solution \eqref{eq:circ_sol_diag}, so e.g.\
\be
\label{eq:sectionsoln}
z_{1,i}&=&   \frac{x_{i,{\rm out}}-x_{i,{\rm in}} }{2\sqrt{2}\cosh \kappa_i \tilde \tau}\cosh (\omega_i\tilde \tau_i + i m_i \sigma_i+ \phi_{1,i})
\nn\\ {\rm with} &&  \tilde \tau_i(x_0) = \frac{1}{\kappa_i}\arctanh \biggsbrk{ \frac{2 x_0 - x_{i,{\rm in}} - x_{i,{\rm out}}}{x_{i,{\rm out}}-x_{i,{\rm in}}} }  \; .
\ee
In particular, this implies
\be
  \mathbf{z}_i^2 = (x_{i,{\rm out}} - x_0) (x_0 - x_{i,{\rm in}}) \; ,
\ee
so that $\mathbf{z}_i=0$, i.e.\ the segments reach the boundary, at times $x_0= x_{i,{\rm in}} $ and $x_0 = x_{i,{\rm out}}$. For simplicity, we will consider in the following the incoming string with $x_{1,{\rm in}}=-a$, while the outgoing strings end at times $x_{2,{\rm out}} =0$ and $x_{3,{\rm out}} = a$. The range of all $\sigma_i$ coordinates is $0$ to $2\pi$. At the intersection time the string is alternatively parametrized by $\sigma_1$ or by $\sigma_2$ and $\sigma_3$, where the interval $\sigma_1 \in [0,2\pi \tfrac{m_2}{m_1}]$ is identified with $\sigma_2\in[0,2\pi]$, and the interval $\sigma_1 \in [2\pi \tfrac{m_2}{m_1},2\pi]$ is identified with $\sigma_3\in[0,2\pi]$.

In the AdS part of the space-time the string solution is the same as for the BMN string discussed in \secref{sec:class-3pt}. Thus, we readily know the position of the intersection point $(x_\nt,z_\nt)$ and the ``virtual'' end-points  $x_{1,{\rm out}}$, $x_{2,{\rm in}}$ and $x_{3,{\rm in}}$. Using $\vec{a}_1 = (-a,0,0,0)$, $\vec{a}_2 = (0,0,0,0)$, and $\vec{a}_3 = (a,0,0,0)$ in \eqref{eqn:intersection-point}, we find
\be
  x_\nt = \frac{\alpha_1 \alpha_2  - \alpha_2 \alpha_3 }{ \alpha_1 \alpha_2 + \alpha_2 \alpha_3 + 4 \alpha_1 \alpha_3 } \, a
  \comma
  z_\nt = \frac{2 \sqrt{ \alpha_1 \alpha_2 \alpha_3 (\alpha_1 + \alpha_2 + \alpha_3) } }{ \alpha_1 \alpha_2 + \alpha_2 \alpha_3 + 4 \alpha_1 \alpha_3 } \, a
  \; .
\ee
Now, the virtual endpoints are determined by the formulas in \eqref{eqn:table} and read
\be
  x_{1,{\rm out}} = \frac{\alpha_2}{\alpha_2+2\alpha_3} \, a
  \comma
  x_{2,{\rm in}} = \frac{\alpha_1+\alpha_3}{\alpha_1-\alpha_3} \, a
  \comma
  x_{3,{\rm in}} = - \frac{\alpha_2}{2 \alpha_1+\alpha_2} \, a
  \; .
\ee

On the sphere the string is extended in two planes. Let us consider string ``1'' in the 1-2 plane and at the intersection time $x_\nt$
\be
  \frac{z_{1,1}+i z_{2,1}}{\mathbf{z}_1}=\frac{1}{\sqrt{2}}{\rm exp}\Big[ \omega_1\tilde \tau_{1,{\rm int}}+\phi_{1,1}+i 
m_1\sigma_1\Big] \; ,
\ee
where $\tilde \tau_{1,\nt} = \tilde \tau_1(x_\nt)$ and where the function $\tilde\tau_1(x_0)$ was defined in \eqref{eq:sectionsoln}. We can determine the constant phases of segments ``2'' and ``3'', $\phi_{1,2}$ and $\phi_{1,3}$, in terms of $\phi_{1,1}$, $x_{2,{\rm in}}$, $x_{3,{\rm in}} $, $x_{1,{\rm out}} $ and $\tau_{\rm int}$ by demanding that these segments overlap with segment ``1''. Specifically, we choose a parametrization so that the point $\sigma_1=0$ on the first string coincides with the point $\sigma_2=0$ on the second and thus we determine
\be
  \phi_{1,2} = -\omega_2\tilde\tau_{2,{\rm int}}+\omega_1 \tilde\tau_{1,{\rm int}}+\phi_{1,1} \; .
\ee
While $0\leq \sigma_2\leq 2 \pi$, the coordinate on the first string runs between $0\leq \sigma_1\leq 2\pi\tfrac{m_2}{m_1}$. Then, taking $\sigma_3=0$ to coincide with $\sigma_1=2\pi\tfrac{m_2}{m_1}$ 
\be
  \phi_{1,3} = -\omega_3\tilde\tau_{3,{\rm int}}+\omega_1 \tilde\tau_{1,{\rm int}}+\phi_{1,1}+ 2i \pi m_2 \; .
\ee
Similarly, we find 
\be
  \phi_{2,2} \eq -\omega_2\tilde\tau_{2,{\rm int}}+\omega_1 \tilde\tau_{1,{\rm int}}+\phi_{2,1} \; , \nln
  \phi_{2,3} \eq -\omega_3\tilde\tau_{3,{\rm int}}+\omega_1 \tilde\tau_{1,{\rm int}}+\phi_{2,1}- 2i \pi m_2 \; .
\ee
Now we are required to minimize the action on the remaining undetermined phases $\phi_{1,1}$ and $\phi_{2,1}$, however, as they appear linearly in the action, they simply give a delta function imposing $J_1-J_2-J_3=0$. 

To this point most of the considerations are independent of the precise form of the vertex operators, only in using the form of the exponentially large AdS terms to perform the minimization have we made concrete use of the explicit form. As this string is a point particle in the AdS space its seems reasonable that it is identical to the point particle string result and further one expects the same delta-functions for the angular momenta regardless of any derivative in the sphere portions of the vertex operator. However we now wish to evaluate the full action, including boundary terms, on the solution and this will be more sensitive to the details of the vertex operator. It is important to note that not only does the vertex operator \eqref{eq:simp_circ_vert} give the correct boundary conditions, but as we will see, with the appropriate definition of $\Delta$ in terms of $J$ and $m$, it gives finite results in a non-trivial fashion when evaluated on the solution. While this may also not be enough to fix the form of the vertex operator, as again any polynomial terms most likely will not change this fact, it does give another constraint.

Inserting the solution into the action, we have from the boundary terms, here using the boundary terms 
${\cal B}^{\rm R}_i\sim -\ln W^{\rm R}_i$
originating from the vertices of the type $V^{\rm R}(\vec{a}_i)=W_i^{\rm R}$, and with ${\cal B}^{\rm R}=\sum_i {\cal B}^{\rm R}_i$,
\be
\label{eq:bdcirc}
e^{- {\cal B}^{\rm R}}&=&\frac{1}{(a+x_{1,{\rm out}} )^{\Delta_1}(x_{2,{\rm in}} )^{\Delta_2}(a-x_{3,{\rm in}} )^{\Delta_3}}
{\rm exp} (-2 \omega_1 J_1\tilde \tau_{1,\nt}-2 \omega_2 J_2\tilde \tau_{2,\nt}-2 \omega_3 J_3\tilde \tau_{3,\nt})
\nn\\
& &
\times \left(\frac{a+x_0}{x_{1,{\rm out}} -x_0}\right)^{\tfrac{\Delta_1}{2}-\tfrac{2J_1^2}{\Delta_1}}_{x_0=-a}
\left(\frac{x_0}{x_{2,{\rm in}} -x_0}\right)^{\tfrac{\Delta_2}{2}-\tfrac{2J_2^2}{\Delta_2}}_{x_0=0}
\left(\frac{a-x_0}{x_0-x_{3,{\rm in}} }\right)^{\tfrac{\Delta_3}{2}-\tfrac{2J_3^2}{\Delta_3}}_{x_0=a}~.
\ee
By itself this contribution is divergent however in this case, unlike for the BMN string, the bulk action is non-vanishing when evaluated on this solution and moreover it cancels against the divergent part of the boundary action. 

Let us consider the first string segment where we find
\be
\label{eq:actcirc}
{\cal S}_1&=&\frac{\sqrt{\lambda}}{2\pi}\int d\sigma_1\int^{\tau_{\rm int} }_{ -\tfrac{i a}{\sqrt{2}} }d\tau ~ {\cal L}_1 
=-\frac{\lambda m_1^2}{2\Delta_1}\ln \left. 
\frac{(a+x^0)}{(x_{1,{\rm out}} -x^0)} \right|^{x^0=x^0_{\rm int} }_{x^0=-a}~.
\ee
This term is also divergent from the $x_0=-a$ singularity at the boundary, however it nicely combines with the divergence from the bulk contribution using $\Delta^2_i=4J_i^2+\lambda m_i^2$. Combining the boundary contributions \eqref{eq:bdcirc}, the exponential of the bulk action contribution, \eqref{eq:actcirc}, including similar terms for the other segments, ${\cal S}=\sum_i {\cal S}_i$, and using the expressions for $x_{1,{\rm out}} , x_{2,{\rm in}} , x_{3,{\rm in}} $ we find that
\be
\label{eq:three_pt_circ_win}
\langle V^{\rm R}_1V^{\rm R}_2V^{\rm R}_3\rangle &=&e^{-{\cal S}- {\cal B}^{\rm R}}\\
&=&\frac{1}{a^{\Delta_1+\Delta_2+\Delta_3} 2^{\Delta_1-\Delta_2+\Delta_3} }
\sqrt{\frac{\alpha_1^{\alpha_1} \, \alpha_2^{\alpha_2} \, \alpha_3^{\alpha_3} \, (\alpha_1+\alpha_2+\alpha_3)^{\alpha_1+\alpha_2+\alpha_3}}{(\alpha_1+\alpha_2)^{\alpha_1+\alpha_2}(\alpha_1+\alpha_3)^{\alpha_1+\alpha_3}(\alpha_2+\alpha_3)^{\alpha_2+\alpha_3}}}~.\nn
\ee
Quite remarkably this is exactly the same answer as for the BMN string with, however, the dimensions $\Delta_i$ being quite different. We note that while the answer is the same, it comes about in a somewhat non-trivial fashion combining terms from the bulk action and the boundary terms. 

The form of the vertex operator and the fact that this result looks so similar to the BMN three-point function suggests that we have rather calculated the correlator of massive point-particle states, or perhaps merely some subleading saddle-point contribution to such a correlator (for a genuine point particle state one would imagine that the 
leading saddle point would be a $\sigma$ independent solution). Indeed, as we have already mentioned, we are not able to identify uniquely the correct vertex operator. However, if the difference is merely due to additional polynomial terms then \eqref{eq:three_pt_circ_win} provides the exponentially large contribution, and the remaining normalization comes from
evaluating the polynomial terms on the solution
\be
\langle V(\tau_1,\vec{a}_1) V(\tau_2,\vec{a}_2) V(\tau_3,\vec{a}_3)\rangle_{\cal S}=\langle U(\tau_1)U(\tau_2)U(\tau_3)\rangle_{{\cal S}+{\cal B}(W^{\rm R}_1,W^{\rm R}_2,W^{\rm R}_3)}~.
\ee
It is interesting to note that even if we take the vertices of the type $V^{\rm B}(\tau_i)$ the factors depending 
on the T-dualized coordinates do not contribute and so the vertex essentially becomes $V^{\rm R}(\tau_i)$.
If we use the relations between the parameters and the charges calculated in \cite{Buchbinder:2010gg}
then we find exactly the same answer, in the leading semiclassical approximation we are working in, as 
\eqref{eq:three_pt_circ_win}. If we used the relations following from using the modified bulk action boundary conditions
the answer will again be the same as long as we use a modified dispersion relation $\Delta=\Delta(J,m)$
to guarantee a finite result. 

\section{Quantum three-point function}
\label{sec:quant-3pt}

In this section we wish to make some comments on the generalization of the considerations of 
\secref{sec:quant-2pt} to the three-point function. Having shown that the fluctuation action for the
bosons and fermions, for extremal and non-extremal correlators, 
is simply that of light-cone gauge fixed plane-wave string theory the result is almost 
immediate.  In essence we wish to outline how from
the light-cone path integral evaluation
of their correlation function one reproduces the cubic
Hamiltonian in plane-wave light-cone string field theory \cite{Spradlin:2002ar, Spradlin:2002rv, Pankiewicz:2002tg}. 

The interest in this rederivation is that we can extend our considerations to non-extremal correlators. 
Moreover, while we do not address these topics in this work it is to be hoped that 
these methods can be more straightforwardly generalized to higher order worldsheet
quantum corrections and to other classical string vacua. Finally, a related point is that it is currently 
not  clear that the prefactor for the supersymmetric vertex operator derived for the 
plane wave geometry is correct when applied to the AdS/CFT correspondence. This stems from the fact
that the full AdS$_5$ $\times$ S$^5$ conserved charges in the plane-wave limit, particularly the supercharges, do
not correspond exactly with the charges calculated directly in the plane-wave geometry when applied to off-shell states
and furthermore the off-shell algebra is not identical \cite{Beisert:2005tm, Beisert:2006qh,Arutyunov:2006ak,Klose:2006zd}. 
Thus an approach which explicitly follows from a perturbative expansion of the full 
AdS$_5$ $\times$ S$^5$ action is useful.

As in the classical case we will consider state ``1" as incoming, so that $\tau_1\to -\infty$, 
and states ``2" and ``3" as outgoing, $\tau_{2,3} \to \infty$.  The string worldsheet is thus composed of 
three segment each corresponding to the resepctive segment of the classical solution. As the fluctuation analysis is local on the worldsheet we can trivially repeat the calculation of \secref{sec:quant-2pt}, thus we find three regions each of which is described by a plane wave action but each with a different mass,  
 \be
 S_{\rm fluc}=\int_{\tau_1}^{\tau_{\rm int}}\kern-3pt  d\tau \kern-3pt \int_0^{l_1 }\kern-3pt  d\sigma~ L_{(1)} 
 +\int_{\tau_{\rm int}}^{\tau_2} \kern-3pt d\tau\kern-3pt \int_{0}^{l_2} \kern-3pt  d\sigma~ L_{(2)}
 +\int_{\tau_{\rm int}}^{\tau_3} \kern-3pt d\tau \kern-3pt  \int_0^{l_3}\kern-3pt  d\sigma~ L_{(3)}
 \ee
 where
 \be
 & &L_{(i)}=\frac{1}{2\pi}\Big[\ \dot X_i^2+{\acute X}_i^2+ \mu_i^2 X(i)^2\Big]~,
 \ee
 with $\mu_i={\cal J}_i$ the mass of the fluctuations of the 
 fields on the three regions of the worldsheet. 
 Here we note that on the different segments  are parameterized such that each segment has worldsheet length $l_i=2\pi$. On each segment 
 we can rescale the spatial coordinate and the
 worldsheet time 
 \be
 \zeta_i=\xi_i+i \eta_i =\alpha_i \tau_i +i  |\alpha_i| \sigma_i
 \ee
 so that each mass is unity but now $l_i=2\pi |\alpha_i|$ with $\alpha_i={\cal J}_i$. In the extremal case one has 
 \be
\alpha_1=\alpha_2+\alpha_3
 \ee
 which can be identified with the conservation of the light-cone momentum. 
Just as for the N-point function we can integrate out the transverse fluctuations by introducing the
Green's function $N(\sigma,\tau;\sigma',\tau')$. 
In terms of the original choice for the worldsheet spatial coordinate 
(where each segment has period $2\pi$) we take as an ansatz for the  
general expansion of the Green's function in terms of the Neumann coefficients $N^{ij}_{m,n}$,
 \be
 N(\sigma,\tau_i;\sigma',\tau_j)&=&-\delta^{ij} \frac{2}{\mu_i}-\delta^{ij} \sum_{n=1}^\infty \frac{4}{\omega_{i,n}} \Big[\cos( n \sigma)\cos( n \sigma')+ \sin( n \sigma)\sin( n \sigma')\Big] \\
 & &+8\sum_{n,m}
  \frac{ e^{-\omega_{i,m} |\tau_i|-\omega_{j,n}|\tau_j|} }{\sqrt{\omega_{i,n}\omega_{j,m}}}
  \Big[N_{m,n}^{ij}\cos( m \sigma)\cos( n \sigma')+ N_{-m,-n}^{ij}\sin( m \sigma)\sin( n \sigma')\Big]\nn
 \ee
 where we have  the individual plane-wave oscillator frequencies $\omega_{i,n}=\sqrt{n^2+\mu^2_i}$.
 Starting from \eqref{eq:npointNeu}, and taking
 the normalization, including the functional determinant, to be one, we can write
 the three point function as
 \be
 \langle V_1 V_2 V_3\rangle= e^{-{\cal S}_{\rm cl}- {\cal B}_{\rm cl}}~ C^{123}
 \ee
 with 
 \be
 C^{123}=  \int \prod_{i,n,I} dP^I_{i,n} \psi_{k_i}(P^I_{i,n})~~e^{\sum_i H_{{\rm lc},i} |\tau_i| +\frac{1}{4}\sum_{i,j}\int d\sigma'd\sigma''~ 
 P^I_i(\sigma')N(\sigma',\tau_i;\sigma'',\tau_j)P^I_j(\sigma'')}~.\nn
 \ee
 Using the oscillator expressions for the wavefunctions, and recalling that we treat string ``1" as 
 incoming, ``2" and ``3" as outgoing, we write the coefficients $C^{123}$ in terms of the Neumann coefficients,
 \be
 C^{123}=\langle \{k_{1,n},k_{2,n},k_{3,n}\}|{\rm exp} \Big[\sum_{\substack{n,m\\i<j}} 
N^{ij}_{m,n}a_{i,m}^{I\dagger}a_{j,n}^{I\dagger}\Big]|0\rangle~.
 \ee
 Exactly parallel considerations for the fermions produce result in a fermionic contribution to the exponential, $\exp\big[\sum Q^{ij}_{m,n}b_{i,m}^{\dagger}b_{j,n}^{\dagger}\big]$, where $Q^{ij}_{nm}$ are the usual fermionic Neumann matrices and our conventions for the oscillators, $b^\dagger_n$, are  defined in 
 \eqref{eq:ferm_osc}. 
 
\paragraph{Neumann coefficients.} We have thus reproduced the three-point amplitude (for the purely 
bosonic theory) in terms of the Neumann coefficients. In flat space they can be most easily 
found by using the conformal invariance of the light-cone gauge fixed theory
whereas in the plane-wave theory their determination is slightly more complicated. They are found
by demanding continuity and conservation of momentum across the string junctions,\footnote{Here we use the rescaled
worldsheet coordinates so that in the extremal case the string worldsheet length is conserved across the interaction.}
\be
\langle X_1(\eta)-X_2(\eta)-X_3(\eta)\rangle=\langle P_1(\eta)+P_2(\eta)+P_3(\eta)\rangle=0~.
\ee
These equations imply that \cite{Spradlin:2002ar, Spradlin:2003xc}
\be
\label{eq:Neumann_def}
N_{mn}^{ij} =\delta^{ij} \delta_{mn} -2\sqrt{\omega_{i,m}\omega_{j,n}}(X^{(i)T}
\Gamma^{-1}X^{(j) })_{mn}
\ee
with $\Gamma_{mn}=\sum_i\sum_l \omega_{i,l}X^{(i)}_{ml}X^{(i)}_{nl}$ and 
\be
X^{(1)}_{mn} = \delta_{mn} \comma
X^{(2)}_{mn} =\frac{1}{\pi}(-1)^{m+n+1}\frac{\sin( \pi m y)}{n-m y} \comma
X^{(3)}_{mn} =\frac{1}{\pi}(-1)^{n}\frac{\sin \pi m (1-y)}{n-m(1- y)}
\ee
with $y=\tfrac{{\cal J}_2}{{\cal J}_1}$. Explicit expression were given in a series of papers \cite{Spradlin:2002ar, He:2002zu} and perhaps most efficiently in \cite{Lucietti:2004wy} hence we will not repeat the derivation here but refer the reader to the references.

\paragraph{Prefactor.} It is long known from the flat space case that superstring amplitudes cannot be simply calculated in light-cone gauge as overlap amplitudes of vertex operators but that non-trivial insertions must be made at the string interaction points \cite{Mandelstam:1974hk, Mandelstam:1974fq, Mandelstam:1985wh}. In the RNS formulation of open superstrings such insertions are schematically of the form $S^i_1\partial X^i$ where $S^i_a$ are the usual Grassmann valued, spacetime vectors. The explicit form of these insertions is determined by demanding Lorentz invariance of the path integral. In the Green-Schwarz formulation of open superstrings in flat space an ansatz for the insertion was proposed by Mandelstam in \cite{Mandelstam:1985wh}: for each joining point we include a  factor of $\sum_I | I  \rangle \partial X^I $ on the single string segment, where the state $| I\rangle$ is a vector state in the supersymmetric formalism. For the closed string the insertion on the single string segment, our segment ``1", at the joining point is a tensor product $\sum_{IJ} |IJ\rangle \partial X^I  \bar \partial X^J $. Using the complex fluctuation fermions introduced in \eqref{eq:real_spinors} we can define the tensor product state to be
\be
 \label{eq:ferm_insertion}
 |IJ \rangle& =& \delta^{IJ}+\frac{1}{2}  \lambda^a   \lambda^b  \gamma^{IJ}_{ab} +\frac{1}{4!}(   \lambda^a   \lambda^b   \lambda^c   \lambda^d) t^{IJ}_{abcd}\nn\\
 &  &+\frac{1}{6!}(    \lambda^c    \lambda^d    \lambda^e   \lambda^f   \lambda^g    \lambda^h) \gamma^{IJ}_{ab}\epsilon^{ab}{}_{cdefgh}+ \frac{1}{8!}(    \lambda^a   \lambda^b   \lambda^c    \lambda^d    \lambda^e   \lambda^f   \lambda^g    \lambda^h) \epsilon_{abcdefgh}~.
 \ee
In the path integral each insertion must be contracted with a term from the boundary, thus for example
\be
\langle (\dot X^I(\sigma_{1})-\acute X^I(\sigma_{1})) \dots\rangle &=&\langle\{k_{1,n},k_{2,n},k_{3,n}|\Big[\sum_{s=1}^3\sum_{n,m} ({\omega_n +n})\cos n \sigma_1 N^{1s}_{n,m}a^{I\dagger}_{1,m}\Big]\dots|0\rangle\nn\\
&\sim& \frac{1}{ 
\sqrt{\sigma_1-\sigma_{1,{\rm int} } }}
\langle\{k_{1,n},k_{2,n},k_{3,n}| \sum_m K_{1,m} a^{I\dagger}_{1,m}\dots|0\rangle
\ee
however this divergent as $\sigma_1\to \sigma_{1,{\rm int}}=0$ and must be regularized. 
 
In flat space light-cone string field theory the complete vertex function, including the prefactor was given in \cite{Green:1982tc, Green:1983hw}. This was generalised to the plane-wave geometry in \cite{Spradlin:2002ar, Spradlin:2002rv}, see also \cite{Pankiewicz:2002gs,Pankiewicz:2002tg,Pankiewicz:2003kj} and in particular the fermionic component is morally similar to  \eqref{eq:ferm_insertion}. The prefactor is constrained by demanding that it is consistent with the plane-wave superalgebra, however this is not sufficient to uniquely determine it and alternative forms of the prefactor were proposed  \cite{Chu:2002eu, Chu:2002wj,DiVecchia:2003yp, Lee:2004cq}. It was shown in \cite{Dobashi:2004nm, Lee:2004cq} that a linear combination of the different prefactors with equal weights is consistent with the supergravity limit of holography. There is also an intrinsic ambiguity whereby the cubic vertex can be modified by making a unitary transformation\footnote{This ambiguity is also present in flat space as one can make similar unitary transformations, however such ambiguities do not contribute to S-matrix elements, the observables in flat space, at least to leading order in perturbation. We thank H.~Shimada for this point.}, for example adding the cubic vertex of \cite{DiVecchia:2003yp} can be seen as such a transformation \cite{Shimada:2004sw}. Currently there does not exist a first principles derivation of the cubic vertex.

In the case at hand a further distinction must be made: as has been explicitly shown 
for the usual BMN string i.e. the Lorentzian analogue of \eqref{eqn:class-sol-2pt} with $c_0=1, b_0=-1$ \cite{Arutyunov:2006ak,Klose:2006zd}, the dynamical supercharge, $Q^-$, to quadratic order in the transverse fields is given by
\be
Q^-=-\sqrt{2}\int d\sigma e^{\tfrac{-i \Pi x^-}{2}} \Big[ 2\pi P^I \gamma^I \lambda- i \acute X^I \gamma^I \bar \lambda- i \mu X^I \gamma^I\Pi \lambda
\Big],~
\ee
 which differs from the plane-wave expressions by the non-local factor, exp${\tfrac{-i \Pi x^-}{2}}$. This has two effects,
 firstly the superalgebra relevant for determining the prefactor is modified 
  \be
 \{Q^-_a,\bar Q^-_b\}=\delta_{ab} H- i \mu (\gamma_{rs}\kern-3pt&& \kern-1pt \Pi)J^{rs}+i \mu (\gamma_{r's'}\Pi)J^{r's'}~,\nn\\
 \{\bar Q_a^-,\bar Q_b^-\}= \delta_{ab} \tilde P~, & & \{ Q_a^-, Q_b^-\}= \delta_{ab} \tilde K~, 
\ee
where $J^{rs}$ and $J^{r's'}$, $r,s=1,\dots,4$ $r',s'=5,\dots, 8$ are the SO$(4)$ rotations and 
the momentum generators $\tilde P$ and $\tilde K$ are not present in the plane-wave algebra but correspond
to the central extensions of the psu$(2|2)$ algebra introduced in the context of the AdS/CFT
duality in \cite{Beisert:2005tm}. Such terms were considered in the calculation of the plane-wave cubic vertex
in \cite{Kishimoto:2010ak}, where under the assumption that $\tilde P$ and $\tilde K$ receive no corrections,
the prefactor was shown to be that found in  \cite{Pankiewicz:2003kj,Lee:2004cq}. 

A second feature is 
that the supercharges act on products of excitations, at least excitations of the same string segment, with a non-trivial coproduct
\cite{Beisert:2005tm,Beisert:2006qh,Arutyunov:2006ak,Klose:2006zd,Arutyunov:2006yd}. On the worldsheet
the definition of the coproduct made essential use of the decompactification limit
of the worldsheet.  It would be very interesting to generalize this coproduct to strings 
with multiple segments and to repeat the calculation of the prefactor. Unfortunately we
do not currently have such a definition, however if we additionally restrict our considerations to excitations with momenta that
are very small compared to the string charges we expect that the plane-wave calculation of
the prefactor is valid. In this limit we can take over all the results from light-cone string field theory.

\section{Conclusions and discussion}

In this work we have considered the light-cone gauge approach to  the 
study of worldsheet correlation functions of vertex operators  for strings in AdS$_5\times$S$^5$. 
For the case of two-point functions, we have shown that the family of euclidean BMN strings provide the saddle-point approximation to the path integral where the boundary conditions are given by the components of the vertex operators that scale
 as $(\dots)^{\sqrt{\lambda}}$. The action, both the bulk and boundary contributions, evaluated on these solutions is completely
 finite, a result due to the fact that the vertex operators describe physical on-shell states
 satisfying the Virasoro constraints. We then analyzed the fluctuations around the saddle point and showed that, as expected, 
 the fluctuations are described by the plane-wave action with the masses depending only on the total charge
 $\Delta$ of the solution. As is expected, the quantum corrections do not effect the space time dependence of the
 correlator, but additionally the dependence on the particular orientation of the solution on the compact S$^5$ drops
 out. Including the fluctuations about the classical solutions we can also define the vertex operators for
 near-BMN states with non-vanishing worldsheet momentum. At the quadratic level it is straightforward 
 to see that the vertex operators do not mix, and we can identify this worldsheet calculation as the holographic 
 two-point function of the gauge theory near-BMN operators.
 
 We then studied the saddle-point calculation of the worldsheet correlator of three BMN string vertices. Following a similar calculation
 in Lorentzian signature \cite{Janik:2010gc}, the saddle point is given by finding the intersection of three euclidean BMN strings. 
 We are able to explicitly determine the coordinates of the intersection point and evaluate the action on the
 solution reproducing the standard space-time dependence for three-point functions in a conformal theory. 
 We consider both extremal correlators, where all three strings rotate in the same plane, and non-extremal correlators
 for strings moving in orthogonal planes and intersecting at a single point. In the first case we find the usual extremal
 relation, $\Delta_1=\Delta_2+\Delta_3$ and the solution degenerates so that the intersection point in fact lies on 
 the boundary. In the non-extremal case, the intersection point is generically in the bulk. We further extend these
 results to circular winding strings, which are described by point-like geodesics in the AdS subspace but are extended along great circles of the
 S$^5$. Here, once again we find a finite result and one that surprisingly has the same form as that for the BMN strings. In this case, we used the proposed vertex of \cite{Ryang:2010bn} in evaluating the correlator, however it is not certain that this is the final correct form. If there are additional polynomial terms, then their contribution, evaluated on the solution, must be included. 
We also consider the vertex operator proposed earlier in \cite{Buchbinder:2010gg} which involved T-dualized angles on the sphere. This vertex 
sourced the same saddle point solution in our formulation after changing the boundary conditions on the bulk action, moreover
the form of the correlation function was essentially identical. 
Regardless of  the ultimately correct form of the vertex operator it should provide the same boundary conditions and thus 
the same classical solution should provide the leading saddle-point contribution to the three-point correlator. 
 
Returning to the BMN strings, it is straightforward to study the quantum corrections to the three-point functions. Following 
the standard light-cone approach to the evaluation of the string path integral and using the earlier fluctuation analysis
we make contact with the results of string field theory and earlier holographic calculations of three-point structure constants. 
As mentioned, the fluctuations only depend on the total charges of the vertices and not their specific orientation. This
implies that, as is to be expected from conformal invariance, the structure constants have no dependence 
on the boundary locations of the vertex operators but also that they do not depend on the relative orientations of
the charges in the compact directions i.e on the ${\bf n}$ vectors. This appears to agree with the results from
the gauge theory. It also suggests that the structure constants for the extremal and non-extremal correlators 
are smoothly related i.e. given the structure constants for a generic non-extremal correlator 
$C^{123}(\Delta_1,\Delta_2,\Delta_3;\{k_i\})$ we can then find the extremal expression by analytic continuation,  as was suggested in \cite{Liu:1999kg} and which is the philosophy taken in recent weak coupling calculations \cite{ Escobedo:2010xs, Escobedo:2011xw}. 
Under this assumption, we make direct contact with the results of light-cone string field theory. The leading quantum corrections to the structure constants are thus found by calculating matrix elements between oscillator states with contractions made using the Neumann matrices and including an appropriate insertion at the intersection point.

An obvious and important open direction is how to include further quantum corrections to the correlation 
functions of near-BMN operators. To this end it may be useful to take a slightly different perspective, one which proved 
useful in the study of the spectral problem, and consider the decompactification limit of all three strings. That is, rescale the worldsheet spatial coordinate on all three segments so that $0\leq \sigma_i\leq 2\pi {\cal J}_i$ and take ${\cal J}_i\to \infty$ while keeping the ratios fixed and then study the worldsheet theory perturbatively in a small momentum expansion. This would involve
including further terms in the expansion of the action which can be found straightforwardly, and which in turn would give rise to
at least three sources of correction,
\begin{itemize}
\item Corrections to the vertex operators: Just as one calculated the corrections to the energies of string states \cite{Callan:2003xr,Callan:2004uv,Frolov:2006cc}, one 
can perturbatively calculate the corrections to the string states themselves. In effect one would need to calculate the
corrections to the two point functions of vertex operators and diagonalize the resulting mixing matrix. 
\item Corrections to the prefactor: To find corrections to the prefactor one would need to repeat the supersymmetry analysis 
of \cite{Pankiewicz:2003kj,Lee:2004cq, Kishimoto:2010ak} but allow for a more general ansatz for invariants
presumably involving more powers of bosonic fields. One may likely need to allow for the non-trivial coproduct action 
for the supercharges. If this is insufficient to fix any ambiguities, it would be very desirable to have a definition of the higher, 
non-local charges  which may fix the prefactor uniquely. In flat space and for the RNS string, an alternative to  insertions at the
interaction points was to introduce ${\cal N}=1$ worldsheet supersymmetry  i.e. supersheets \cite{Berkovits:1985ji, Berkovits:1987gp}. 
A similar  result was shown for flat space Green-Schwarz strings in \cite{Berkovits:1991qg}
and for plane waves in \cite{Berkovits:2002vn}, whether this
can be repeated for the full AdS$_5\times$S$^5$ case remains an open question. 
\item Corrections to the worldsheet propagator and correspondingly the Neumann matrices.
\end{itemize}
Regarding the last point, for the two-point function it is known that, in the decompactification limit, the exact propagator for a single magnon is found by the replacement $\omega_p=\sqrt{1+p^2}\to \sqrt{1+4 \sin^2(p/2)}$. As the Neumann matrices are determined simply by the mode expansion and continuity it is tempting to conjecture, and so we shall, that a similar replacement will produce the correct, all-order Neumann matrices via the usual relation
\be
N_{mn}^{ij} =\delta^{ij} \delta_{mn} -2\sqrt{\omega_{i,m}\omega_{j,n}}(X^{(i)T}
\Gamma^{-1}X^{(j) })_{mn}~,
\ee
but using the exact dispersion relations. It would be interesting to ask if such equations can be solved, along the lines of \cite{Lucietti:2004wy} in terms of generalized $\mu$-deformed Gamma-functions
but again with $\omega_{p}=\sqrt{1+4 \sin^2(p/2)}$ and $p_n=n/L$. In any case, it should be possible to determine, at least perturbatively, the corrections by including higher order terms from the worldsheet action. 

A related direction is to study whether the method of patching together two-point classical string solutions to
find three-point solutions can be generalized to a wider range of  configurations. In this work we considered the simplest
circular winding strings on the sphere and on the sphere it should be straightforward to consider more general strings, with
more general angular momenta, etc. 
Whether the same can be done for strings extended in the AdS space, for example folded spinning strings, remains to be 
seen. Similarly it would be worthwhile to calculate the  quantum corrections to more general 
configurations, even for the simplest circular string. Here finding the fluctuation action and calculating 
the corrections to the non-excited vacua should be straightforward though understanding the definition of the
vertex operator, particularly the explicit form of the $U_{\Delta, J;m}(\vec \sigma)$ function, becomes essential. 

As mentioned our considerations are always for Euclidean worldsheet signature and Euclidean AdS, corresponding to the calculation of Euclidean correlation functions in the boundary theory. The proposals of \cite{Janik:2010gc} included using physical strings with Lorentzian worldsheets to holographically calculate correlators in Minkowski space-time, which requires finding  classical solutions describing the joining and splitting of physical strings. Such classical solutions have been found in \cite{Peeters:2004pt,Peeters:2005pb, Casteill:2007td,Murchikova:2011ea} and a complete general solution on the $\mathbb{R}\times S^3$ subspace has recently been given in \cite{Vicedo:2011vn}. Our calculation differs not only in using Euclidean worldsheets but also in the construction of the saddle points. While we demand that the string segments overlap we do not separately demand that the time-derivatives also match. Rather, we determine the remain parameters by minimizing the action upon varying the intersection point. Nonetheless it would be very interesting to see if similar methods can be used in the Euclidean theory, in particular those making use of the worldsheet integrability, something which has not played an overt role in our considerations.

\section*{Acknowledgements}

We are grateful to N. Beisert, S. Frolov,  R. Janik, H. Shimada, and particularly to J. Minahan and  A. Tseytlin  for useful conversations. 

\appendix

\section{Fermionic fluctuation action}
\label{app:fermions_fluc}

For the most part we have neglected to include the fermions however even at the level of quadractic fluctuations they play a crucial role. Here we will analyze the quadratic fermionic fluctuations about the classical solution \eqref{eqn:class-sol-2pt}.\footnote{As for the bosonic fluctuations, we focus on the solution corresponding to the two-point function
with 
 ${\bf n}_1=-{\bf n}^\ast_2={\bf n}$. As we will see, this is not a significant assumption as
 the fluctuation spectrum depends only on the overall charge $\Delta$ and not the boundary
 position or plane of rotation.} 
 As previously discussed the light-cone gauge action is equivalent to using the diagonal gauge fixed Lagrangian where the length of the worldsheet is determine by the light-cone momentum.
Thus  we start from the action of \cite{Giombi:2009gd} for the fermions $\theta_i$ and $\eta_i$, $i=1,2,3,4$, and their conjugates $\theta^i=(\theta_i)^{\dagger}$, $\eta^i=(\eta_i)^{\dagger}$,
\footnote{Due to our specific gauge fixing, worldsheet time coordinates are rescaled by inverse powers of $s_{\rm cl}$ compared to \cite{Giombi:2009gd}.} which to quadratic order in the fermions is, 
\be
 \Lagr_{\rm ferm} \eq  i(\theta^i\dot\theta_i +\theta_i\dot\theta^i +\eta^i\dot\eta_i +\eta_i\dot\eta^i)
  +2i \frac{\dot z^M z^N}{z^2} \, \eta_i (\bar{\rho}^{MN})^i{}_j \eta^j \nln & &
  +2i \frac{z^M}{s_{\rm cl}z^3} \eta^i(\rho^M)_{ij}       \acute{\theta}^j
  +2i \frac{z^M}{s_{\rm cl} z^3} \eta_i(\bar{\rho}^M)^{ij} \acute{\theta}_j \; .
\ee
The $4\times4$ matrices $\rho^M = (\rho^M)_{ij}$ are the off-diagonal blocks of the Dirac matrices in six dimensions in chiral representation, we will not need explicit expression but a convenient representation can be found in \cite{Giombi:2009gd}. We have also defined $\bar{\rho}^M = (\rho^M)^{ij} = (\rho^M_{ij})^\dagger$, and
\be
  \bar{\rho}^{MN} = (\bar{\rho}^{MN})^i{}_j = \Half (\bar{\rho}^M \rho^N - \bar{\rho}^N \rho^M)
  \comma
  \rho^{MN} = (\rho^{MN})_i{}^j = \Half (\rho^M \bar{\rho}^N - \rho^N \bar{\rho}^M)~.
\ee
These matrices satisfy the Clifford algebra
\be
  &&
  \bar{\rho}^M \rho^N + \bar{\rho}^N \rho^M = 2 \delta^{MN} 
  \comma
  \rho^M \bar{\rho}^N + \rho^M \bar{\rho}^N = 2 \delta^{MN} 
  \; .
\ee
The matrix part of the expressions we are dealing with in this section is always an alternating product of some number of factors $\rho^M$ and $\bar{\rho}^N$ whose $\grSO(6)$ indices are contracted either to $\mathbf{n}$ or $\mathbf{n}^*$. Therefore, it is useful to define the notation $(\mathbf{n}\cdot\rho) = n^M \rho^M$. Using the properties of $\mathbf{n}$ given in \eqref{eqn:properties-n} and the Clifford algebra, it follows that
\be
  (\mathbf{n}\cdot\rho)(\mathbf{n}\cdot\bar{\rho}) = (\mathbf{n}^*\cdot\rho)(\mathbf{n}^*\cdot\bar{\rho}) = 0
  \comma
  (\mathbf{n}\cdot\rho)(\mathbf{n}^*\cdot\bar{\rho}) + (\mathbf{n}^*\cdot\rho)(\mathbf{n}\cdot\bar{\rho}) = 4
  \; ,
\ee
and the same formulas with $\rho\leftrightarrow\bar{\rho}$ exchanged. 
%
We now substitute the classical solution into the action. In order to make the formulas more compact, we introduce an angle $\alpha$ by
\be
  e^{ i\alpha} = \frac{(c_0-x_0)e^{\phi} }{\sqrt{(x_0-b_0)(c_0-x_0)}} \comma
  e^{-i\alpha} = \frac{(x_0-b_0)e^{-\phi}}{\sqrt{(x_0-b_0)(c_0-x_0)}} \; ,
\ee
and the function $F(\tau)=(x_0-b_0)(c_0-x_0)$, which allows us to write
\be
  \mathbf{z}_\cl \eq \half \sqrt{F} \lrbrk{ e^{-i\alpha} \mathbf{n}^* + e^{i\alpha} \mathbf{n} } \; .
\ee
The action becomes
\be
 \Lagr_{\rm ferm} \eq i (\theta^i\dot\theta_i +\theta_i\dot \theta^i +\eta^i\dot\eta_i +\eta_i\dot\eta^i)
 + \frac{c_0-b_0}{\sqrt{2} F}~\eta_i (n^{*M} n^N \rho^{MN})^i{}_j\eta^j\\
 &&
 +\frac{i}{F s_{\rm cl}} \eta^i\lrsbrk{ (\mathbf{n}^*\cdot\rho) e^{-i\alpha} + (\mathbf{n}\cdot\rho) e^{i\alpha} }_{ij} \acute{\theta}^j
 +\frac{i}{F s_{\rm cl}} \eta_i\lrsbrk{ (\mathbf{n}^*\cdot\bar{\rho}) e^{-i\alpha} + (\mathbf{n}\cdot\bar{\rho}) e^{i\alpha} }^{ij} \acute{ \theta}_j \; .\nn
\ee
We can remove the explicit time dependence by rotating the fermions. To this end, we define the matrices
\be
  R = R^i{}_j = \Half n^{*M} n^N \bar{\rho}^{MN}
  \comma
  S = S^i{}_j = \cos \frac{\alpha}{2} - i R \sin \frac{\alpha}{2} \; ,
\ee
which satisfy
\be
  R^\dagger = R
  \comma
  R^2 = \unit
  \comma
  S^\dagger = S^{-1}
  \; .
\ee
Using these rotation matrices, we redefine the fermions as
\be
  \theta^i = S^i{}_j \tilde{\theta}^j
  \comma
  \eta^i = S^i{}_j \tilde \eta^j
  \; ,
\ee
and the same for $\eta$. By virtue of the identities $S^\dagger R S = R$ and
\begin{align}
  S^\trans (\mathbf{n}\cdot\rho) S & = e^{-i\alpha} (\mathbf{n}\cdot\rho) \; , &
  S^\dagger (\mathbf{n}\cdot\bar{\rho}) S^{\dagger\trans} & = e^{-i\alpha} (\mathbf{n}\cdot\bar{\rho}) \; , \\
  S^\trans (\mathbf{n}^*\cdot\rho) S & = e^{i\alpha} (\mathbf{n}^*\cdot\rho) \; , &
  S^\dagger (\mathbf{n}^*\cdot\bar{\rho}) S^{\dagger\trans} & = e^{i\alpha} (\mathbf{n}^*\cdot\bar{\rho}) \; ,
\end{align}
the $\alpha(\tau)$ dependence of the last three term in the Lagrangian disappears. However, the redefinition of the fermions introduces extra contibutions from the time derivative terms. Using
\be
  S^\dagger \dot{S} = - \frac{i\dot{\alpha}}{2} R
  \comma
  \dot{\alpha} = \frac{c_0-b_0}{\sqrt{2}F(\tau)} \; ,
\ee
we find
\be
  \theta_i \dot{\theta}^i \eq \tilde{\theta}_i S^{\dagger i}{}_j \partial_\tau (S^j{}_k \tilde{\theta}^k)
  =\tilde{\theta}_i \dot{\tilde{\theta}}^i - i\frac{c_0-b_0}{2\sqrt{2}F(\tau)} \, \tilde{\theta}_i R^i{}_j \tilde{\theta}^j \; ,
\ee
and similarly for the other kinetic terms. Combining all contributions, we have
\be
 \Lagr_{\rm ferm} \eq \frac{1}{s_{\rm cl} F(\tau)} \Big[ i s_{\rm cl}F(\tau) (
     \tilde{\theta}^i\dot{\tilde{\theta}}_i
   + \tilde{\theta}_i\dot{\tilde{\theta}}^i
   + \tilde{\eta}^i \dot{\tilde{\eta}}_i
   + \tilde{\eta}_i \dot{\tilde{\eta}}^i )
 + i \tilde{\eta}^i (\mathbf{n}^* + \mathbf{n})\cdot\rho_{ij} \acute{\tilde{\theta}}^j
 + i \tilde{\eta}_i (\mathbf{n}^* + \mathbf{n})\cdot\bar{\rho}^{ij} \acute{\tilde{\theta}}_j \nln
 && \hspace{10mm}
 + \frac{s_{\rm cl}(c_0-b_0)}{\sqrt{2}} \, \tilde{\theta}_i R^i{}_j \tilde{\theta}^j
 + 3 \frac{s_{\rm cl}(c_0-b_0)}{\sqrt{2}} \, \tilde{\eta}_i R^i{}_j \tilde{\eta}^j \Big] \; .
\ee
Finally we can redefine the worldsheet time in the same manner as for the bosonic fluctuations $d\tilde \tau=d\tau/(s_\cl F(\tau))$. We thus find the plane-wave Lagrangian for eight complex fermions with masses depending on the solution parameters, 
\be
  s_\cl \frac{(c_0-b_0)}{\sqrt{2}} = i \frac{\Delta}{\sqrt{\lambda}}
\ee
We can redefine fermions once more
\be
  \tilde{\zeta}_i = \Half \tilde{\eta}^j (\mathbf{n}^* + \mathbf{n})\cdot\rho_{ji}
\ee
so that
\be
  \tilde{\zeta}_i \dot{\tilde{\zeta}}^i = \tilde{\eta}^i \dot{\tilde{\eta}}_i
  \comma
  \tilde{\zeta}_i R^i{}_j \tilde{\zeta}^j = -\tilde{\eta}_i R^i{}_j \tilde{\eta}^j
\ee
and then make a final time dependent rotation $\tilde \theta^i=(e^{-i \omega R\tilde \tau})^i{}_j\hat \theta^j$
and $\tilde \zeta^i=(e^{- i \omega R\tilde \tau})^i{}_j\hat \zeta^j${\ }\footnote{As usual there is an issue with reality conditions for the 
fermions, here we assume that $\omega \tilde \tau$ is real.}.
    Choosing 
   $\omega=s_{\rm cl}(c_0-b_0)/2\sqrt{2}$, and dropping hats and tildes we have,
   \be
 \Lagr_{\rm ferm} \eq
  i ({\theta}^i\dot{{\theta}}_i
   + {\theta}_i\dot{{\theta}}^i
   + {\zeta}^i \dot{{\zeta}}_i
   + {\zeta}_i \dot{{\zeta}}^i )
 + 2i ({\zeta}_i \acute{{\theta}}^i
    -  {\zeta}^i \acute{{\theta}}_i )
 + 2i \frac{\Delta}{\sqrt{\lambda}} \, {\theta}_i R^i{}_j {\theta}^j
 - 2 i \frac{\Delta}{\sqrt{\lambda}} \, {\zeta}_i R^i{}_j {\zeta}^j \; 
   \ee
which is the action in terms of two 4-component complex spinors.
We note that the hermitian matrix $R$ has eigenvalues $2\times +1$ and $2\times -1$. We can thus
bring it to the form $\tilde \Pi={\rm diag}(1,1,-1,-1)$ by a unitary transformation on the fields $\theta$ and 
$\zeta$
\footnote{ As both of these fields have the same 
mass matrix they are both diagonalised by the same transformation.}.
We can introduce $\vartheta^a$, $a=1,\dots, 8$, with 
$\vartheta=\tfrac{1}{\sqrt{2}}({ \vartheta}^1-i  { \vartheta}^2)$, $\bar \vartheta=\tfrac{1}{\sqrt{2}}({ \vartheta}^1+i  { \vartheta}^2)$ 
where
\be
\label{eq:real_spinors}
\vartheta^1=  \matr{c}{\theta_i + \theta^i \\ \zeta_i + \zeta^i} \comma \vartheta^2=\matr{c}{i(\zeta_i - \zeta^i) \\ -i(\theta_i - \theta^i)} \; ,
\ee
so that we can rewrite  the action in terms of an 8-component complex spinor
\be
 {\cal L}_{\rm ferm}=\frac{i}{2}(\bar \vartheta \dot { \vartheta} +\vartheta\dot{ \bar \vartheta})+\frac{i}{2}(\bar \vartheta \acute {\bar \vartheta} -\vartheta \acute \vartheta )- i {\bar \vartheta} M \vartheta~,
\ee
 where the mass matrix is symmetric, block off-diagonal with $(\Delta/\sqrt{\lambda})\tilde \Pi$ on the off-diagonals. We can make a further change of basis so that $M=(\Delta/\sqrt{\lambda})\Pi$ with $\Pi={\rm diag}(\unit_4,-\unit_4)$. In terms of SO$(8)$ gamma matrices, with an appropriate choice of representation, $\Pi=\gamma_1\gamma_2\gamma_3\gamma_4$. This
 is simply the plane-wave fermionic action of \cite{Metsaev:2001bj}. 
 
\paragraph{Oscillator expansion.} The oscillator expansion of the fermionic field $\vartheta$
 and its conjugate momentum $\lambda=\tfrac{1}{2\pi}\bar \vartheta$ is given by 
 \be
 \vartheta&=&\vartheta_0+\sqrt{2}\sum_{n=1}^\infty(\vartheta_n \cos n \sigma +\vartheta_{-n} \sin n \sigma)~,\nn\\
\lambda  &=&\frac{1}{2\pi}\Big[\lambda_0+\sqrt{2}\sum_{n=1}^\infty(\lambda_n \cos n \sigma +\lambda_{-n} \sin n \sigma)\Big]~.
 \ee
In terms of BMN creation and annihilation operators we have
\be
\label{eq:ferm_osc}
\vartheta_n = \frac{1}{2\sqrt{\mu}}&&\kern-5pt\left( A_n b_n +  B_n b_{-n}^\dagger\right)~,~~{\rm with}\\
A_n=\frac{1}{\sqrt{\omega_n}} (\sqrt{\omega_n- n}+\Pi \sqrt{\omega_n+n})~,&&
B_n=\frac{i}{\sqrt{\omega_n}} (-\sqrt{\omega_n+n}+\Pi \sqrt{\omega_n-n})\nn
\ee
 \section{A toy model for non-extremal correlators}
 \label{app:toy_non_ext}
Although we do not have a general description for the intersection of solutions corresponding to generic non-extremal
 correlators we  can consider an appropriate ansatz/toy model. In essence, we consider a general point particle ansatz, 
 however rather than solving the equations of motion we simply insert this ansatz into 
 into the action and drop the $\sigma$ dependence. We can expect this to capture the leading semiclassical approximation 
 for point-like BMN
 strings but  
  nonetheless  we are only looking at a toy model. 
 
 As we have discussed in the main text, the BMN geodesics we are interested in correspond to straight line trajectories 
 in Euclidean six-dimensional space. The path integral approach to point particles in spherically symmetric potentials is well studied, see \cite{Kleinert:2004ev}, and making the usual 
 change to spherical coordinates, while slightly subtle, is naturally useful. For a point particle of mass $M$, moving in $D$ flat dimensions, with coordinate vector $\mathbf{z}$, we can expand the path integral calculation of a quantum mechanical amplitude in ultra-spherical harmonics $Y_{l,{\bf m}}({\bf \hat z})$ with ${\bf \hat z}$
 a $D$-dimensional unit vector, $z^2=\mathbf{z}^2$, ${ \hat{\bf  z}}=\tfrac{\mathbf{z}}{z^2}$, i.e.
 \be
 \langle \mathbf{z}_b, \tau_b|  \mathbf{z}_a, \tau_a\rangle =\frac{1}{(z_a z_b)^{(D-1)/2}}
 \sum_{l=0}^\infty \langle { z}_b,\tau_b|{ z}_a,\tau_a\rangle_l \sum_{\bf m} Y_{l{\bf m}}({\bf \hat z}_b) Y_{l{\bf m}}({\bf \hat z}_a)^{\ast}
 \ee
 where the radial amplitude is defined by
 \be
  \langle { z}_b,\tau_b|{ z}_a,\tau_a\rangle_l=\int^{{ z}={ z}_b}_{{ z}=\mathbf{z}_a} {\cal D}{ z}~ e^{-S_{{\rm radial}, l}}
 \ee
 with the effective radial action being
 \be
 S_{{\rm radial}, l}=\frac{M}{2} \int d\tau~ \big[\dot { z}^2+ ``\frac{1}{M^2}\frac{(l+D/2-1)^2-1/4}{{ z}^2}"\Big]
 \ee
 and where the quotation marks imply that one cannot naively consider this action in the path integral when $z$ goes to zero
 but rather one must perform the time slicing of the Cartesian action and then change variables. In effect this changes
 the numerical constant in the numerator however for our 
 immediate semiclassical considerations  this is not relevant and  results in a subleading
 correction.   
 
 Returning to the light-cone worldsheet theory, we consider the point particle limit and perform the above expansion
 for  the six dimensional space spanned by $z^M$, $M=1,\dots, 6$. Defining the shorthand for the radial partial amplitude from 
 the i-th boundary to the intersection point
 \be
  \langle \{x_i,\bar x_i, { z}_i\}, \tau_i|\{x_{\rm int},\bar x_{\rm int}, { z}_{\rm int}\}, \tau_{\rm int}\rangle_{l_i}=\langle i|{\rm int}\rangle_{l_i}~,
  \ee
 we can now make the same expansion for the path integral but include the vertex operators at the boundary. 
The correlator for incoming particle ``1" and outgoing particles ``2" and ``3" is 
\be
\langle V^{(3)}V^{(2)}V^{(1)}\rangle
&=&\int \prod_{i=1}^3 dX(i)^I \int dX^I_{{\rm int}} 
e^{B^{(3)}} e^{B^{(2)}{}} e^{B^{(1)}{}^\ast}\sum_{l_1, l_2, l_3}  \langle 3|{\rm int}\rangle_{l_3} \langle2|{\rm int}\rangle_{l_2}
\langle1|{\rm int}\rangle_{l_1}^\ast \nn\\
& &~~
\sum_{{\bf m}_1, {\bf m}_2, {\bf m}_3}
Y_{l_3 {\bf m}_3} ({\bf \hat z}_3) Y_{l_3 {\bf m}_3}({\bf \hat z}_{\rm int})^\ast~
Y_{l_2 {\bf m}_2}({\bf \hat z}_{2}) Y_{l_2 {\bf m}_2}({\bf \hat z}_{\rm int})^\ast~
Y_{l_1 {\bf m}_1}({\bf \hat z}_{\rm int}) Y_{l_1 {\bf m}_1}({\bf \hat z}_{1})^\ast~.\nn
\ee
For the boundary terms coming from the vertex operators we take the AdS part 
to be given by the usual expression while for the sphere part we take a specific ultra-spherical harmonic
\eqref{eq:vertex_harmonics}, 
\be
B^{(1)}&=&B^{(1)}_{\rm radial}+B^{(1)}_{\rm sphere}\nn\\
&=&
\Delta^{(1)} \ln \left(\frac{{ z_{1}}}{{ z_{1}}^2+(\vec x_{1}- \vec a_{1})^2}\right)+\ln Y_{l'_{1},{\bf m}'_{1}}({\bf \hat z}_{1})
\ee 
and outgoing 
\be
B^{(2,3)}=\Delta^{(2,3)} \ln \left(\frac{{ z_{2,3}}}{{ z_{2,3}}^2+(\vec x_{2,3}- \vec a_{2,3})^2}\right)+\ln Y^\ast_{l'_{2,3},{\bf m}'_{2,3}}({\bf \hat z}_{2,3})~.
\ee 
Using the orthogonality of the ultra-spherical harmonics 
\be
\int d^{5}{\bf \hat z}~ Y^\ast_{l, {\bf m}}(\bf \hat z) Y^\ast_{l', {\bf m}'}(\bf \hat z) =\delta_{ll'}\delta^{(4)}_{{\bf m},{\bf m'}}
\ee
we have 
\be
\langle V^{(3)}V^{(2)}V^{(1)}\rangle
&=&\int \prod_{i=1}^3 d^2x_i  dz_i \int d^2x_{\tau_{\rm int}} dz_{{\rm int}}
e^{B^{(3)}_{\rm radial}} e^{B^{(2) }_{\rm radial }} e^{B^{(1)\ast }_{\rm radial}}
  \langle 3|{\rm int}\rangle_{l_3} \langle2|{\rm int}\rangle_{l_2} \langle1|{\rm int}\rangle_{l_1}^\ast \nn\\
& &~~~\times
\int d^5 
{\bf \hat z}_{\rm int} ~Y^\ast_{l'_3, {\bf m}'_3 }
  ( {\bf \hat z}_{\rm int} )~
 Y_{l'_2 {\bf m}'_2}({\bf \hat z}_{2})~
 Y_{l'_1 {\bf m}'_1}({\bf \hat z}_{1})~.
\ee
The radial component of the path integral can be evaluated using the saddle-point approximation and the
earlier expressions for the BMN geodesic in the AdS$_5$ subspace; this will reproduce the standard space-time dependence. The angular part corresponds to the three-point function structure constants  and  thus we see that generically the non-extremal correlator is proportional to 
\be
\int d^5 
{\bf \hat z}_{\rm int} ~Y^\ast_{l_3, {\bf m}_3 }
  ( {\bf \hat z}_{\rm int} )~
 Y_{l_2 {\bf m}_2}({\bf \hat z}_{\rm int})~
 Y_{l_1 {\bf m}_1}({\bf \hat z}_{\rm int})
 \ee
 which is as expected see e.g. appendix B of \cite{Lee:1998bxa}. 
 We should emphasize again that here we are only treating the point particle part of the path integral
 and we should really find the appropriate classical solutions to the full path integral. 
%

%

\bibliographystyle{JHEP}
\bibliography{3pt}
 
\end{document}